\documentclass[conference]{IEEEtran}

\usepackage{xcolor}
\usepackage{cite}
\usepackage{amsmath,amssymb,amsfonts}
\usepackage{algorithmic}
\usepackage{graphicx}
\usepackage{textcomp}
\usepackage{xcolor}
\usepackage[hyphens]{url}
\usepackage{fancyhdr}
\usepackage{hyperref}
\usepackage[capitalise,noabbrev,nameinlink]{cleveref}
\usepackage{comment}
\usepackage{tcolorbox}
\usepackage{multirow}
\usepackage{soul}
\usepackage[inkscapelatex=false]{svg}
\usepackage{xfrac}
\usepackage{cite}
\usepackage{amsmath,amssymb,amsfonts}
\usepackage{algorithmic}
\usepackage{graphicx}
\usepackage{textcomp}
\usepackage{xcolor}
\usepackage[hyphens]{url}
\usepackage{fancyhdr}

\newcommand{\ignore}[1]{}

\newcommand{\redd}[1]{ {\bf{#1}}}

\newcommand{\rev}[1]{\textcolor{black}{#1}}

\newcommand{\anishTODO}[1]{\textcolor{red}{#1}}

\definecolor{aliceblue}{rgb}{0.94, 0.97, 1.0}
\usepackage{xcolor}
\usepackage{tcolorbox}
\tcbset{width=1\columnwidth,boxrule=1pt,colback=aliceblue,arc=1pt,left=2pt,right=2pt,boxsep=0.5pt}

\newcommand{\alfa}{$\alpha$ }

\title{ImPress: Securing DRAM Against Data-Disturbance Errors via Implicit Row-Press Mitigation \vspace{-0.1 in}}
\author{\IEEEauthorblockN{Moinuddin Qureshi}
\IEEEauthorblockA{
\textit{Georgia Tech}\\
moin@gatech.edu }
\vspace{-0.2 in}
\and
\IEEEauthorblockN{Anish Saxena}
\IEEEauthorblockA{
\textit{Georgia Tech}\\
anish.saxena@cc.gatech.edu}
\vspace{-0.2 in}
\and
\IEEEauthorblockN{Aamer Jaleel}
\IEEEauthorblockA{
\textit{NVIDIA}\\
ajaleel@nvidia.com}
\vspace{-0.2 in}
}

\usepackage{tikz}

\begin{document}
\maketitle
\pagestyle{plain}

\begin{abstract}

DRAM cells are susceptible to \textit{Data-Disturbance Errors
(DDE)}, which can be exploited by an attacker to compromise
system security. \textit{Rowhammer} is a well-known DDE vulnerability
that occurs when a row is repeatedly activated. Rowhammer
can be mitigated by tracking aggressor rows \textit{inside DRAM (in-
DRAM)} or at the \textit{Memory Controller (MC)}. \textit{Row-Press (RP)} is a
new DDE vulnerability that occurs when a row is kept open for
a long time. RP significantly reduces the number of activations
required to induce an error, thus breaking existing RH solutions.

Prior work on \textit{Explicit} Row-Press mitigation, \textit{ExPress}, requires
the memory controller to limit the maximum row-open-time, and
redesign existing Rowhammer solutions with reduced Rowhammer
threshold. Unfortunately, ExPress incurs significant performance
and storage overheads, and being a memory controller-based
solution, it is incompatible with in-DRAM trackers.



In this paper, we propose \textit{Implicit Row-Press mitigation (Im-
Press)}, which does not restrict row-open-time, is compatible with
memory controller-based and in-DRAM solutions and does
not reduce the tolerated Rowhammer threshold. ImPress treats a
row open for a specified time as \textit{equivalent} to an activation.
We design ImPress by developing a \textit{Unified Charge-Loss Model},
which combines the net effect of both Rowhammer and Row-Press for arbitrary patterns. We analyze both controller-based
(Graphene and PARA) and in-DRAM trackers (Mithril and
MINT). We show that ImPress makes Rowhammer solutions resilient to Row-Press transparently, without affecting the
Rowhammer threshold.

\end{abstract}

\ignore{
DDE
RH what, problem, TRH, solutions (MC-based and In-DRAM)
RP what it is?
Impact of RP (Agnositc). Figure-1 overview.

Express: what it is, adapt existing.  shortcoming.
Our goal.
Impress, insight. Figure 1d
Unified DDM
Impress-D
ImPress-P
We evaluate four trackers.
Key results
Contributions.

}


\section{Introduction}

Relentless scaling over the last four decades has increased the capacity of  DRAM chips from a few megabits to several tens of gigabits. As DRAM cells get smaller, they become prone to inter-cell interference, where the activity in one cell can disturb the data in another cell, leading to {\em Data-Disturbance Errors (DDE)}. DDEs are not just a reliability concern but also a serious security threat, as attackers can exploit DDEs to compromise system security~\cite{seaborn2015exploiting, gruss2016rhjs}.

\vspace{0.05 in}

\noindent{\bf Rowhammer:}  The most well-known DDE vulnerability of DRAM is {\em Rowhammer (RH)}\cite{kim2014flipping}.  Rowhammer occurs when an {\em aggressor} row is activated a large number of times, which causes bit-flips in the neighboring {\em victim} rows. Several studies\cite{seaborn2015exploiting, frigo2020trrespass, gruss2018another, aweke2016anvil, cojocar2019eccploit, gruss2016rhjs, vanderveen2016drammer}  have shown that Rowhammer can be exploited to compromise security. For example, an attacker can flip bits in page-tables to escalate privilege~\cite{seaborn2015exploiting}, flip bits in instruction opcode to bypass authentication~\cite{razavi2016flip}, or analyze flipped bits to infer the data of nearby pages~\cite{kwong2020rambleed}.


The number of activations (ACTs) to the aggressor row required to induce a bit-flip is called the {\em Rowhammer Threshold (TRH)}. The latest publicly available characterization data reports a TRH of 4.8K~\cite{kim2020revisitingRH}. Typical hardware-based defenses for Rowhammer rely on a tracking mechanism~\cite{kim2014architectural,kim2014flipping,PROHIT,kim2014architectural,kim2022mithril,ProTRR,qureshi2022hydra,park2020graphene} to identify aggressors and refresh the victim rows~\cite{hassan2021UTRR}. The tracking can be either at the {\em Memory-Controller (MC)} or within the DRAM {\em (in-DRAM)}.  Solutions for mitigating RH are designed for a {\em specific} TRH, which assumes DRAM will not incur bit-flips if the activation count is below the specified TRH. 
These solutions can be broken if a vulnerability causes bit flips with fewer than TRH activations.

\begin{figure*}[!htb]
    \centering
\includegraphics[width=6.95in]{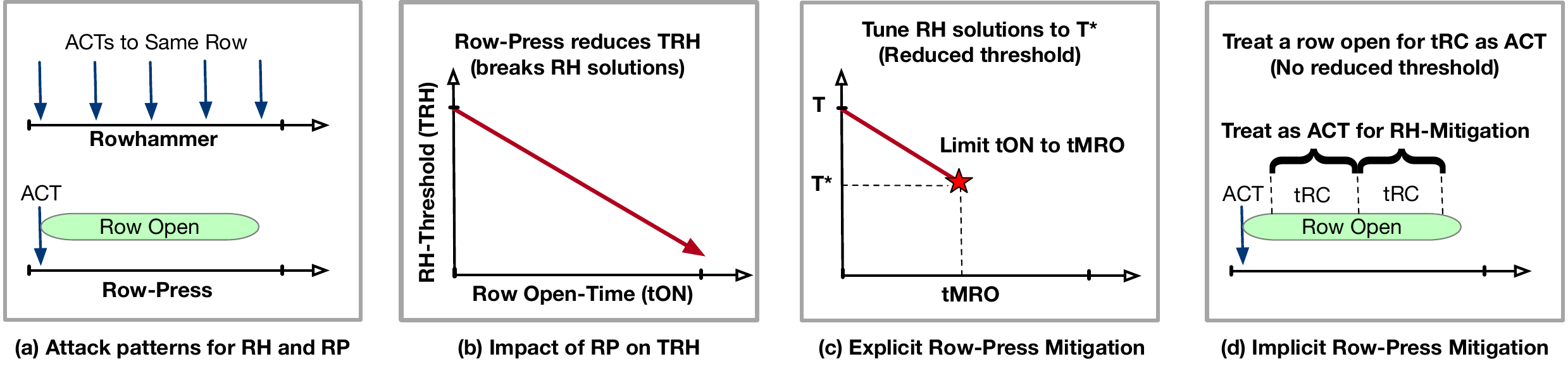}
   \vspace{-0.1 in}
    \caption{Towards practical Row-Press solution: (a) Attack pattern for Rowhammer and Row-Press (b) Impact of RowPress on the Rowhammer Threshold (c) Explicit Row-Press (ExPress)~\cite{rowpress} mitigation, which limits the aggressor row-open time (tON) to tMRO, reduces the tolerated RH threshold  (d) Our proposal, Implicit Row-Press (ImPress) mitigation, treats a row-open for tRC as equivalent to an activation (ACT) for RH-mitigation, retains the same RH threshold.} 
  \vspace{-0.15 in}
    \label{fig:intro}
\end{figure*}

\vspace{0.05 in}
\noindent{\bf Row-Press:} A recent paper~\cite{rowpress} discloses a new DDE vulnerability, {\em Row-Press (RP)}, which occurs when a row is kept open for a long time.  While the row is open, the cells of the neighboring rows {\em slowly} leak charge on the bit-lines. The cumulative charge loss increases with time. Therefore, a Row-Press pattern keeps the row open for as long as possible. The row may eventually close due to a row conflict or refresh operation. Such a Row-Press attack pattern is repeated until the charge on the neighboring cell is depleted enough to cause a flip.  Figure~\ref{fig:intro} (a) compares the pattern of RH and RP.


\vspace{0.05 in}
\noindent{\bf Impact of Row-Press:} The impact of Row-Press depends on how long the row is kept open. Each round incurs an activation of the given row.  Luo et al.~\cite{rowpress} provide a detailed characterization of Row-Press and show that the number of activation rounds required to succeed is 18x to 160x lower compared to the number of activations required by a standalone RH attack. If the row is kept open for 30ms (note that this is not possible in DDR specifications, as they require that refresh be performed within a few tens of microseconds), then a single round of Row-Press attack may be enough to flip a bit. 

Figure~\ref{fig:intro} (b) captures the impact of Row-Press on TRH. RP reduces the number of activations required to cause a bit-flip to much lower than TRH (e.g. 18x compared to Rowhammer alone~\cite{rowpress}). Thus, RP breaks RH-mitigation designed to tolerate a threshold of TRH, as such solutions inherently assume that no bit-flip occurs if the row gets fewer than TRH activations. Therefore, RP is a serious security vulnerability.

\vspace{0.05 in}

\noindent{\bf Explicit Row-Press Management: } 
Luo et al.~\cite{rowpress} also proposed a design to tolerate Row-Press attacks, which forces the Memory Controller (MC) to limit the amount of time a row can be kept open to {\em Maximum Row Open time (tMRO)}. For example, let $TRH$ denote the threshold for the standalone RH attack. 
The number of activations, $T*$, required for Row-Press to flip bits is characterized, with the maximum aggressor open time (tON) being limited to tMRO.
The proposal redesigns the RH-mitigation to cater to the lower threshold, T*, instead of TRH. We term this design {\em Explicit Row-Press (ExPress) Mitigation}. Figure~\ref{fig:intro}(c) provides an overview of ExPress.

\newpage
\noindent{\bf Pitfalls of ExPress: } The key shortcoming of ExPress is that it reduces the tolerated threshold from TRH to T*. Additionally, ExPress suffers from the following three problems: 

(1) High performance overheads: Early row closure reduces the row buffer hits for workloads with good spatial locality.  Furthermore, tuning the RH solution to a lower threshold (T*) increases the rate of mitigation and the associated penalty.

(2) High storage overheads: If the tracking mechanism is based on counters, the number of tracking entries increases due to the reduction in threshold from TRH to T*.  

(3) Incompatibility with in-DRAM Trackers: ExPress is a memory controller-based solution, as it must limit tON to tMRO. Therefore, it is incompatible with in-DRAM Rowhammer schemes that are unaware of tMRO value, unless JEDEC specifications are revised to standardize tMRO.

\vspace{0.05 in}
\noindent{\bf Our Goal:} The goal of our paper is a Row-Press solution which (i) does not place any limit on row-open time, (ii) does not affect the TRH tolerated by a Rowhammer solution, (iii) incurs low performance and storage overheads, and (iv) is applicable to both memory controller-based and in-DRAM RH solutions (without requiring changes to JEDEC specifications).  

\vspace{0.05 in}
\noindent{\bf Our Insight: } Our key insight is that secure RH-mitigation schemes are designed to handle the case when an activation occurs at every tRC (row-cycle time).  If a row is open for a particular time period (say, tRC), then we treat it as {\em equivalent} to causing an activation, and the row participates in the RH-mitigation. Doing so, converts the Row-Press attack into an equivalent Rowhammer attack and lets the RH solution transparently handle RP, without limiting the row open time.

\vspace{0.05 in}
\noindent{\bf Our Solution: } We propose {\em Implicit Row-Press (ImPress) Mitigation}, as shown in Figure~\ref{fig:intro}(d). To drive the design of ImPress, we first develop a {\em Unified Charge-Leakage Model} that combines the effect of both Rowhammer and Row-Press into a single metric. Both RP and RH {\em damage} the data in the cell by causing charge loss, albeit at a different rate.  Our model normalizes the rate-of-damage caused by RP (per tRC) to the rate-of-damage caused by RH (per tRC). Our model estimates the combined damage caused by RH and RP, for any pattern.  

Our first design, {\em Impress-N (Naive)}, operates on integer values of damage, to demonstrate the impact of imprecise damage estimation.  Impress-N divides the time interval between refresh into windows of tRC.  If an activation occurs in the given tRC window, that row participates in RH-tracking.  If a row is open for the full tRC window, then that row is treated as equivalent to causing an activation, and also participates in RH-tracking. Impress-N limits the impact of unmitigated RP to at-most one tRC. 
As Impress-N can underestimate Row-Press activity, it can reduce the threshold (T*) by 1.35x-2x, which is identical to ExPress at the corresponding tON. While ImPress-N has similar impact on threshold, performance, and storage as ExPress, ImPress-N is compatible with in-DRAM tracking as it does not limit row open time. 



Our optimized design, {\em Impress-P (Precise)}, operates on precise values of damage, including non-integer values. ImPress-P dynamically tracks the row-open time (tON) and converts it into an  {\em Equivalent Activation Count (EACT)}. We modify the RH trackers to operate on non-integer values. For example, a probabilistic solution that mitigates with probability {\em p} would now select the row with probability $p \times EACT$. A counter-based tracker would increment the counter by $EACT$ instead of 1. As Impress-P is precise, it maintains the same tolerated threshold (TRH) as a system that does not have RP mitigation.

We analyze ImPress with both memory controller-based (Graphene and PARA) and in-DRAM (Mithril and MINT) trackers, and show that ImPress is effective at tolerating Row-Press for both categories. The storage required for ImPress-P is 1.25x, whereas it is 2x for both ExPress and ImPress-N.

\vspace{0.05 in}
\noindent{\bf Contributions: } This paper makes the following contributions:

\begin{enumerate}
    \item We observe that Row-Press can be transparently handled, without limiting tON, by treating a row open for tRC as equivalent to an activation for RH schemes. 
    
 \vspace{0.05 in}
    \item We develop a {\em Unified Charge-Leakage Model} to capture the net effect of RP and RH for any given pattern into a single number and use this metric to guide our design.
\vspace{0.05 in}
    \item We propose {\em ImPress-N}, which treats a row open for the full time window of tRC as equivalent to an activation. ImPress-N limits the impact of unmitigated Row-Press to tRC, reducing Rowhammer threshold by 1.35x-2x.

\vspace{0.05 in}
    \item We propose {\em ImPress-P}, which tracks tON and uses the damage due to RP and RH precisely into RH solutions. ImPress-P does {\em not} reduce the tolerated RH threshold and tolerates RP with negligible overheads. 
\end{enumerate}

\ignore{
Threat
DRAM
Rowhammer (defenses -- all four)
RowPress (diagram of work and reduction)
Impact of RowPress (Angostic)
}

\newpage
\section{Background and Motivation}

\subsection{Threat Model}

We assume a threat model where the attacker can issue memory requests for arbitrary addresses. The attacker is free to choose the memory system policy (e.g. open-page versus closed-page) that is best suited for the attack. The attacker knows the defense algorithm, including which row has been selected for mitigation. We declare an attack to be successful when it causes a bit-flip at any location in memory. 

\subsection{DRAM: Operation and Timings}

DRAM chips are organized as banks, which are two-dimensional arrays consisting of rows and columns. To access data from DRAM, the memory controller must first issue an activation (ACT) to open the row. The row can continue to be open until it is (a) proactively closed by the memory controller (e.g. closed-page policy) (b) closed due to a row conflict to service data from another row, or (c) closed to perform refresh.

DRAM has deterministic timings, which are specified as part of the JEDEC standards (see Table~\ref{table:Params}).  All data in DRAM is refreshed every tREFW. To reduce the latency impact of refresh, memory is divided into 8192 groups, and a refresh pulse is sent every tREFI interval to refresh one group. DDR5 specifications allow the postponement of up-to 4 refreshes, so the time between refresh can be up to 5 times tREFI.

\ignore{
DDR5 standards support two new commands to support mitigation for Rowhammer. First, {\em RFM (Refresh Management}, which allows the MC to provide time to the DRAM chip to internally perform mitigation. Second, {\em DRFM (Directed Refresh Management)}, where the MC provides the row address to be mitigated to the DRAM chip. The latency of RFM is tRFM and the latency of DRFM is tDRFM.  \cref{table:Params} shows the DDR5 parameters. The two critical parameters for our study are: (1) The maximum number of activations possible within tREFI is 79 and (2) The device performs one Rowhammer mitigation at each refresh event.
}

\begin{table}[!htb]
  \centering
  \vspace{-0.1in}
  \caption{DRAM Timings}
  \vspace{-0.1in}
  \begin{footnotesize}
  \label{table:Params}
  \begin{tabular}{lcc}
    \hline
    \textbf{Parameter} & \textbf{Description} & \textbf{Value} \\ \hline

    tACT     & Time for performing ACT & 12 ns \\ 
    tPRE      & Time to precharge an open row & 12 ns \\ 
    tRAS     & Minimum time a row must be kept open & 36 ns \\ 
     tRC       & Time between successive ACTs to a bank & 48 ns \\ \hline 
        
    tREFW     & Refresh Period & 32 ms \\ 
    tREFI     & Time between successive REF Commands & 3900 ns  \\ 
    tRFC      & Execution Time for REF Command & 350 ns  \\ 

\hline 

\rev{tON}          & \rev{Time the current row is open (dynamic  value)} & \rev{--} \\ 
\rev{tONMax}      &  \rev{Max time a row can be kept open per DDR5} & \rev{19.5 $\mu$s} \\ 
\rev{tMRO}         & \rev{Max time a row can be kept open by the MC} & \rev{--} \\

\hline
  \end{tabular}
  \end{footnotesize}
\end{table}

DRAM systems are susceptible to {\em Data-Disturbance Errors (DDE)}, whereby the operations on one DRAM cell can corrupt the data stored in a nearby cell. An attacker could exploit DDE to compromise system security~\cite{seaborn2015exploiting, frigo2020trrespass, gruss2018another, aweke2016anvil, cojocar2019eccploit, gruss2016rhjs, vanderveen2016drammer, kwong2020rambleed}.  In this paper, we focus on two specific modalities of DDE for DRAM: {\em Rowhammer (RH)} and {\em Row-Press (RP)}.

\subsection{Rowhammer: Problem and Solutions}
Rowhammer~\cite{kim2014flipping} occurs when a row (aggressor) is activated frequently, causing bit-flips in nearby rows (victim). The number of activations to an aggressor row to cause a bit-flip in a victim row is called the {\em Rowhammer threshold (TRH)}. Solutions for mitigating RH must ensure that the victim rows get refreshed before the aggressor row incurs TRH activations.

Typical solutions for mitigating RH rely on a {\em tracking} mechanism to identify aggressor rows, and performing a mitigative refresh on the victim rows. Aggressor-row identification could be done either using {\em activation-counters}~\cite{kim2022mithril,ProTRR,qureshi2022hydra,park2020graphene, saxena2024start, olgun2024abacus} or {\em probabilistically}\cite{kim2014architectural,kim2014flipping,PROHIT,kim2014architectural}.

The tracking can be done at the {\em Memory-Controller (MC)} or transparently inside the DRAM chip {\em (in-DRAM)}. The advantage of in-DRAM tracking is the potential to solve the RH problem inside of the DRAM, without relying on other parts of the system. DDR5 provides support for in-DRAM tracking with {\em Refresh Management (RFM)}. Without loss of generality, we analyze the following four trackers in our study.

\vspace{0.05 in}
\noindent{{\bf Graphene}~\cite{park2020graphene} (Counters, MC-Based):} Graphene uses {\em Misra-Gries} algorithm to identify rows that reach TRH activations and issue a mitigation. The number of tracking entries (per bank) is inversely proportional to the threshold.

\vspace{0.05 in}
\noindent{{\bf PARA}~\cite{kim2014flipping} (Probabilistic, MC-Based):} PARA selects each activation for mitigation with a probability {\em p}, which is determined based on a target failure rate.

\vspace{0.05 in}
\noindent{{\bf Mithril}~\cite{kim2022mithril} (Counters, in-DRAM):} Mithril uses {\em Counter-based Summary} to identify heavily activated row. Mitigation is performed on receiving the RFM command (sent by MC every {\em RFMTH} activations) for the row with the highest count. The number of entries depends on RFMTH and TRH.

\vspace{0.05 in}
\noindent{{\bf MINT}~\cite{MINT} (Probabilistic, in-DRAM):} MINT is our concurrent work that achieves secure mitigation with just a single entry per bank. At each RFM, MINT mitigates the identified aggressor row, and randomly selects which activation slot in the upcoming RFMTH activations will be selected for mitigation. 


\vspace{0.05 in}

These trackers are designed to provide secure RH tolerance for a specific TRH.  However, if the attacker can cause bit-flips in fewer than TRH activations, then the attacker can break all of these designs.  A new vulnerability makes this possible.

\subsection{Row-Press: Bit-Flips With Fewer Activations}

\begin{figure}[!htb]
    \centering
\includegraphics[width=2.75 in]{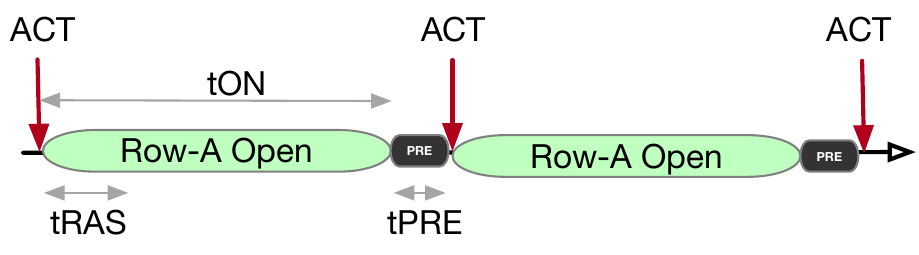}
   \vspace{-0.1 in}
    \caption{Pattern for the Row-Press Attack (PRE denotes Precharge operation)} 
    \label{fig:rp}
\end{figure}

{\em Row-Press (RP)}~\cite{rowpress} is a new DRAM DDE vulnerability, which occurs when a row is kept open for a long time.  When the row is open, the cells of the neighboring rows leak charge on the bit-lines at a non-negligible rate. Over a long time, the total charge loss due to this leakage can become substantial.  Figure~\ref{fig:rp} shows the access pattern for  RP. \rev{Let $tON$ be the time a row is kept open.}  With RP, we keep the aggressor open for a time which is much larger than tRAS. This pattern is repeated continuously until a bit-flip occurs. 



Luo et al.~\cite{rowpress} characterized RP for DDR4 devices and showed that, RP can reduce the number of activations required to induce a bit-flip by 18x (on average, when the row is kept open for one tREFI, which is 7800ns in DDR4) to 156x (on average, when the row is kept open for 9 tREFI, which is 70 $\mu$s for DDR4) compared to standalone RH attacks. As RP can perform bit-flips in much fewer than TRH activations, it can break  RH-mitigations designed for a threshold of TRH.

\begin{figure*}[!tbh]
    \centering
    \includegraphics[width=0.9\textwidth]{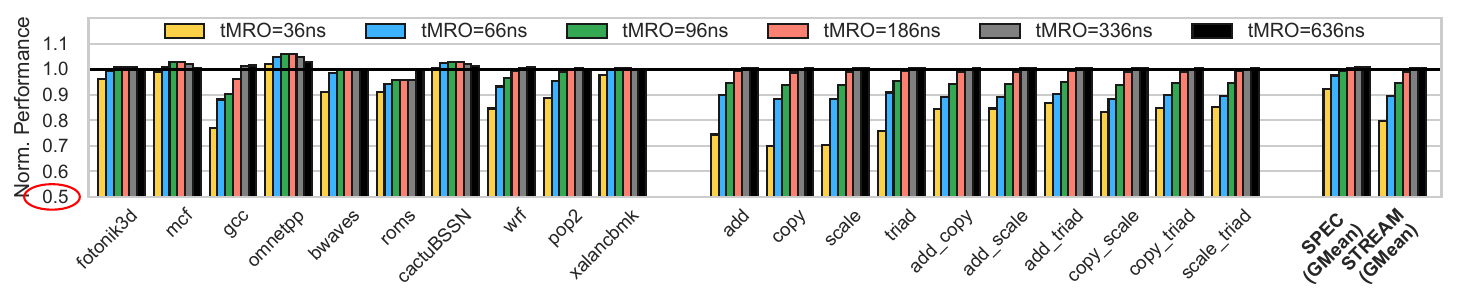}
    \vspace{-0.15in}
    \caption{\rev{Performance impact of limiting the time a row is open to a particular value, termed as tMRO (Maximum Row-Open Time). While SPEC workloads (low/medium spatial locality) are less sensitive to tMRO value, Stream workloads (high spatial locality) can suffer significant slowdown at low tMRO.} }
    \label{fig:ton}
   \vspace{-0.15 in}
\end{figure*}

\newpage

\subsection{Tolerating RP by Limiting the Row-Open Time}

Row-Press exploits the fact that an opened row can continue to be open for a long time, and without any row-conflicts, this time gets constrained only by the time between refresh operations (tREFI, 3900ns for DDR5 and 7800ns for DDR4, although it can be extended with refresh postponement to 5 times tREFI in DDR5 and 9 times tREFI in DDR4).  

\begin{figure}[!htb]
    \centering
\includegraphics[width=3.4 in]{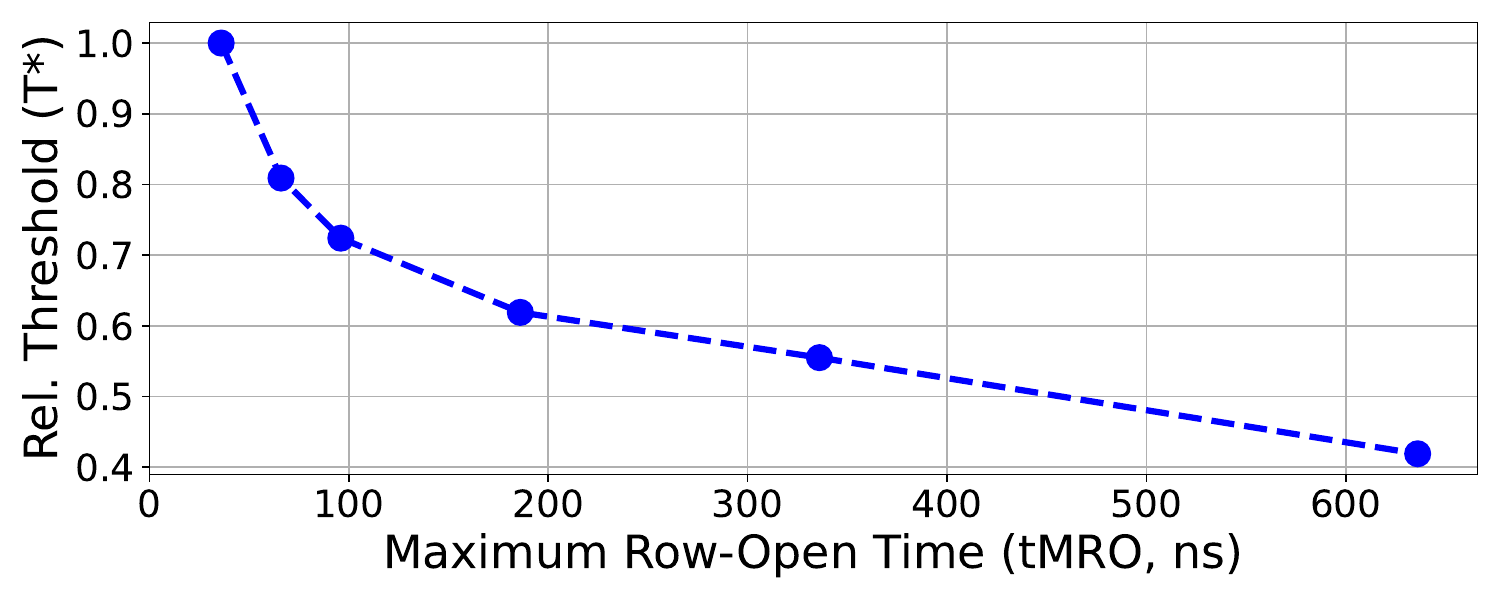}
  \vspace{-0.05 in}
    \caption{Reduction in Tolerated TRH (T*) if the maximum tON is constrained to tMRO (Note: the data is obtained from Table-8 of ~\cite{rowpress})} 
      \vspace{-0.1in}
    \label{fig:tonTRH}
\end{figure}

Luo et al.~\cite{rowpress} propose a solution to mitigate RP by \rev{using the Memory-Controller to limit the {\em Maximum Row-Open Time (tMRO)}.} Figure~\ref{fig:tonTRH} shows the number of activations required on the aggressor row with RP (i.e. change in TRH) as the tMRO is varied from 36ns (minimum value, tRAS) to 630ns.  For example, if tON is limited to 186ns, the {\em effective} threshold (T*) reduces to 62\%.  The RH-mitigation can be redesigned to tolerate this new threshold (T*). As this approach explicitly uses RP to change the threshold of existing algorithms, we term this solution {\em Explicit Row-Press (ExPress) Mitigation}. 

Limiting the time a row can be open can reduce the row buffer hit-rate due to premature closing of the row.  The performance impact of early row closure depends on the type of workload. If the workload has poor row-buffer locality, early closure will not cause a slowdown (it may cause slight improvement due to the removal of precharge from the critical path). However, if the workload has good spatial locality then early row-closure can have a significant performance impact.  Figure~\ref{fig:ton} shows the normalized performance of our system for two classes of workloads: SPEC2017 and Stream, as the tMRO is varied from 36ns to 636ns.  On average, for SPEC, low tMRO has a negligible performance impact, whereas for Stream, low tMRO can cause a significant slowdown (e.g. on average, 10\% for tMRO of 66ns). Thus, ExPress can cause significant slowdowns for an entire category of applications.

ExPress causes additional slowdown as the design must cater to a lower threshold (T*), thus sending more mitigative refreshes. We analyze ExPress for our four trackers.  

\begin{figure}[!htb]
    \centering
    \vspace{-0.1 in}
    \includegraphics[width=0.9\columnwidth]{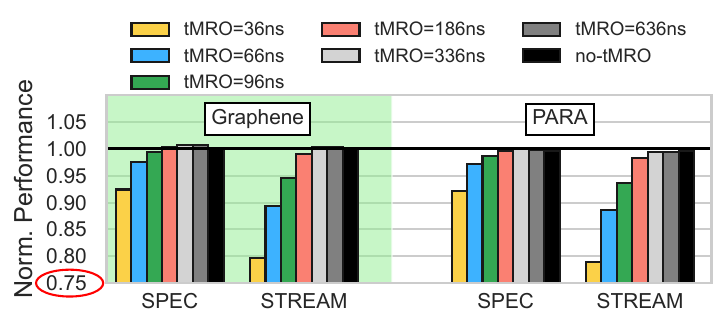}
   \vspace{-0.1 in}
    \caption{Performance of Graphene and PARA as tMRO is varied. Stream has slowdown at low tMRO. [Note: All values are geometric means.] } 
\vspace{-0.15 in}
    \label{fig:tonPARA}
\end{figure}

\vspace{0.05 in}
\noindent{\bf{Graphene:}} Figure~\ref{fig:tonPARA} shows the performance of Express as tMRO is varied. For Stream, the slowdown is significant at low tMRO (due to reduced row-buffer hits). \rev{Furthermore, a higher tMRO increases the effective threshold (T*). So, Graphene must be targetted to an even lower threshold, hence it would need more entries.} At tMRO of 80ns, the storage of Graphene increases from 115KB to 160KB per channel.

\vspace{0.05 in}
\noindent{\bf {PARA:}} Figure~\ref{fig:tonPARA} also shows the relative performance of PARA when the tMRO is varied. The trend is similar to Graphene - negligible impact for SPEC workloads but significant slowdown for Stream workloads at low values of tMRO.

\vspace{0.05 in}
\noindent{ {\bf MINT and Mithril:}} ExPress uses the MC to limit tON to tMRO. \rev{As the tMRO value is decided by the MC, this value is not visible to the trackers inside the DRAM chip (current JEDEC standards do not allow such communication).} Thus, ExPress is incompatible with in-DRAM trackers, so these trackers will continue to be vulnerable to RP. To make ExPress viable for in-DRAM tracking, JEDEC will need to include a new parameter (tMRO). Unfortunately, such changes are hard to incorporate in JEDEC, as all memory/processor vendors will be forced to adhere to new specifications.  Furthermore, any such specification is likely to select the tMRO conservatively.


\subsection{Goal of Our paper}

An ideal solution should tolerate Row-Press transparently, without limiting tON (thus avoiding changes to JEDEC and letting systems choose what performs best for their workloads), should not lower the effective threshold, should have only a minor impact on performance and storage overheads, and should be compatible with both MC-based and in-DRAM designs.  The goal of our paper is to develop such a solution.  
\clearpage
\section{Experimental Methodology}\label{sec:analytical_methods}

\subsection{Performance Methodology}

We use ChampSim~\cite{gober2022championship}, a cycle-level multi-core simulator, interfaced with DRAMSim3~\cite{li:dramsim3}, a detailed memory system simulator. We enhanced DRAMSim3 to support DDR5. Table~\ref{table:system_config} shows the configuration for our baseline system.  We use a {\em Minimalist Open-Page (MOP)} memory mapping with 8 consecutive lines per row.  For RFM, we assume a latency of 205ns (half of tRFC) and use a default RFMTH of 80.

\begin {table}[h!]
\begin{footnotesize}
\begin{center} 
\caption{Baseline System Configuration}
\vspace{-0.1 in}

\begin{tabular}{|c|c|}
\hline
  Out-of-Order Cores           & 8 cores at 4GHz       \\
  Width, ROB size           & 6-wide, 352       \\
  Last Level Cache (Shared)    & 16MB, 16-Way, 64B lines, SRRIP \\ \hline
  Memory size                  & 64GB -- DDR5 \\
 Channels                      & 2 (32GB DIMM per channel) \\  
  Banks x Ranks x Sub-Channels & 32$\times$1$\times$2 \\ 
   Memory-Mapping              & Minimalist Open Page (8 lines)\\  
\hline
\end{tabular}
\label{table:system_config}
\vspace{-0.1 in}
\end{center}
\end{footnotesize}
\end{table}

We use two categories of workloads: First, the 10 SPEC2017~\cite{SPEC2017} (8-core rate mode) traces available from ChampSim to study the impact of tMRO on conventional workloads. Second, 4 streaming workloads~\cite{mccalpin:memory} (8-core rate mode) and 6 mixed streaming workloads (two with 4 copies each), to study the impact of tMRO on high-locality workloads. For each workload, the trace represents the region-of-interest.  We warm-up for 50 million instructions and run each workload for 200 million instructions. We report performance as normalized weighted-speedup. 

\subsection{Reliability Methodology for RH Trackers}

We perform mitigation by refreshing the victim rows. To securely mitigate RH and RP, the parameters of the underlying RH-mitigation scheme must be configured properly. We use a default TRH of 4K~\cite{kim2020revisitingRH}, and show sensitivity in Section~\ref{sec:TRH}.  For probabilistic schemes, we use a target bank-failure rate of 0.1 FIT (1 failure per 10 billion hours, about 30x lower than the rate of naturally occurring errors~\cite{ddr4errors}).

Based on our target failure rate, we configure PARA with p=1/184.  For Graphene, the number of entries is inversely proportional to TRH.  To tolerate a TRH of 4K, Graphene needs 448 entries per bank (115KB SRAM per channel).

Mithril performs mitigation transparently under the RFM command, which is issued every RFMTH activations per bank.  For mitigation, Mithril selects the aggressor row with the highest counter value.  For a given mitigation rate (1 per RFMTH), we determine the number of entries required to tolerate a given threshold using Theorem-1 of~\cite{kim2022mithril}. For example, for RFMTH of 80, Mithril needs 383 entries per-bank (86 KB SRAM per channel) to tolerate a TRH of 4K. 


MINT requires a single-entry per bank to keep track of the row to be mitigated at RFM.  At each RFM, MINT mitigates the given aggressor row and then randomly selects which activation slot in the upcoming RFMTH (e.g. 80) activations will be selected for mitigation at the next RFM.   As MINT lacks configurability (for a fixed RFMTH), we report the threshold tolerated by MINT as the figure of merit.

\newpage
\section{Unified Charge-Loss Model}

To mitigate Row-Press transparently and at low-cost, we propose {\em Implicit Row-Press (ImPress) mitigation}.  ImPress  converts the time incurred in doing Row-Press to an equivalent activation count for Rowhammer. To design ImPress, we first develop a unified charge-loss model for RH and RP.

\subsection{Relative Charge-Loss Model for Rowhammer}

Consider a DRAM cell that is the target of a RH attack. After TRH activations to the aggressor row, the total charge loss suffered by the cell must be above some {\em critical} value to cause a bit flip. We need a model to quantify the {\em Total Charge Loss} incurred after $K$ activations. To keep our model simple, we quantify charge-loss as a {\em relative} metric. Let the {\em relative charge-loss per activation ($C_A$) }be 1 unit. The total charge loss ($TCL_{RH}$) after $K$ activations is given by Equation~\ref{eq:tclrh}.

\begin{equation}
\vspace{0.05 in}
    TCL_{RH} = K \cdot C_A = K \cdot 1 = K
    \vspace{0.05 in}
    \label{eq:tclrh}
\end{equation}

\begin{figure}[!htb]
    \centering
    \vspace{-0.05 in}
\includegraphics[width=2 in]{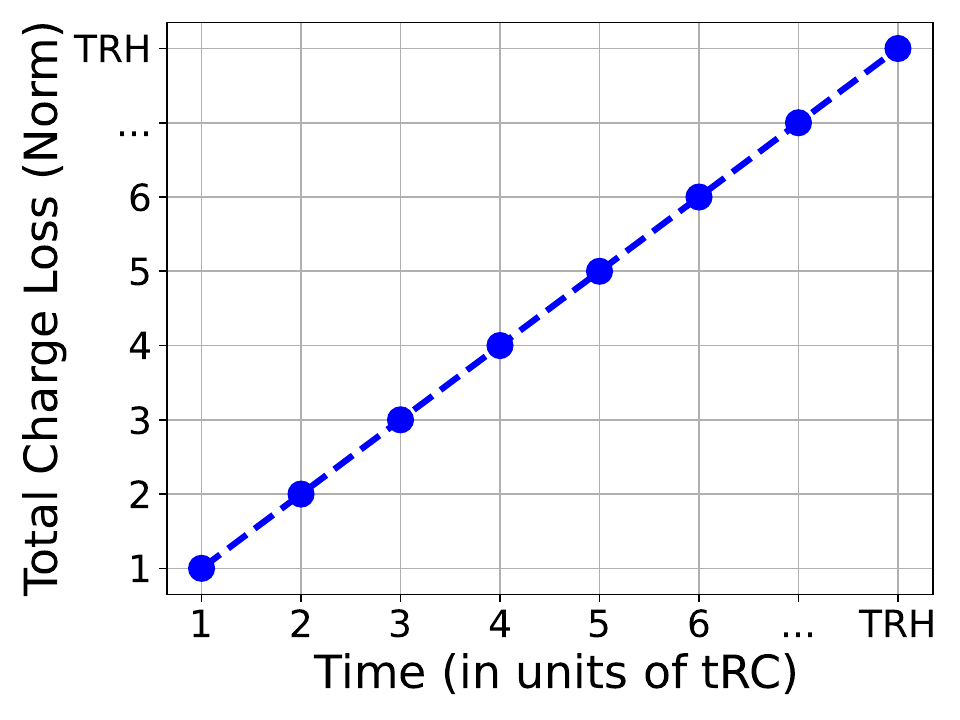}
\vspace{-0.05 in}
    \caption{Relative Charge-Loss Model for Rowhammer} 
    \label{fig:rhmodel}
\end{figure}

As a bit-flip occurs after TRH activations, the total charge loss is $TRH$ units, which represents the value of the {\em critical} charge loss. Figure~\ref{fig:rhmodel} shows the charge-loss model for RH. Note that the time is counted in terms of tRC.  RH is a {\em perfect linear attack} -- one unit of damage in one unit of time.

\subsection{Relative Charge-Loss Model for Row-Press}

\ignore{
The total time to perform one round of the RP attack ($tRPA$) also include the precharge time (tPRE), as each round ends in a precharge. Therefore, 

\begin{equation}
    t_{RPA} =  tON + tPRE
\end{equation}
}

The charge loss for RP comes from two sources: (1) the activation and the time incurred in the first tRC, the impact of which is identical to an RH pattern, so this time period incurs a charge-loss of 1 unit (2) the time-dependent charge loss that occurs because the row is kept open for an {\em additional time} ($tON-tRAS$). As we normalize all times to tRC, we also normalize the {\em additional time} to tRC. The total charge loss ($TCL_{RPA}$) from an RP pattern that keeps a row open for $tON$ time is given by Equation~\ref{eq:rpa}.  

\ignore{
Let $t_{nao}$ be the additional open time {\em normalized} to tRC.

\begin{equation}
    t_{nao} =  \frac{tON-tRAS}{tRC} 
\end{equation}
}

\begin{equation}
    TCL_{RPA} = 1 + f( \frac{tON-tRAS}{tRC} ) 
    \label{eq:rpa}
\vspace{0.1 in}
\end{equation}

Where the function $f$ captures the {\em rate of charge leakage} per unit time (in terms of tRC) for RP. This function can be estimated using the characterization data, or picked conservatively such that it can never be below the observed data.

\begin{figure*}[!tbh]
    \centering
\includegraphics[width=2in]{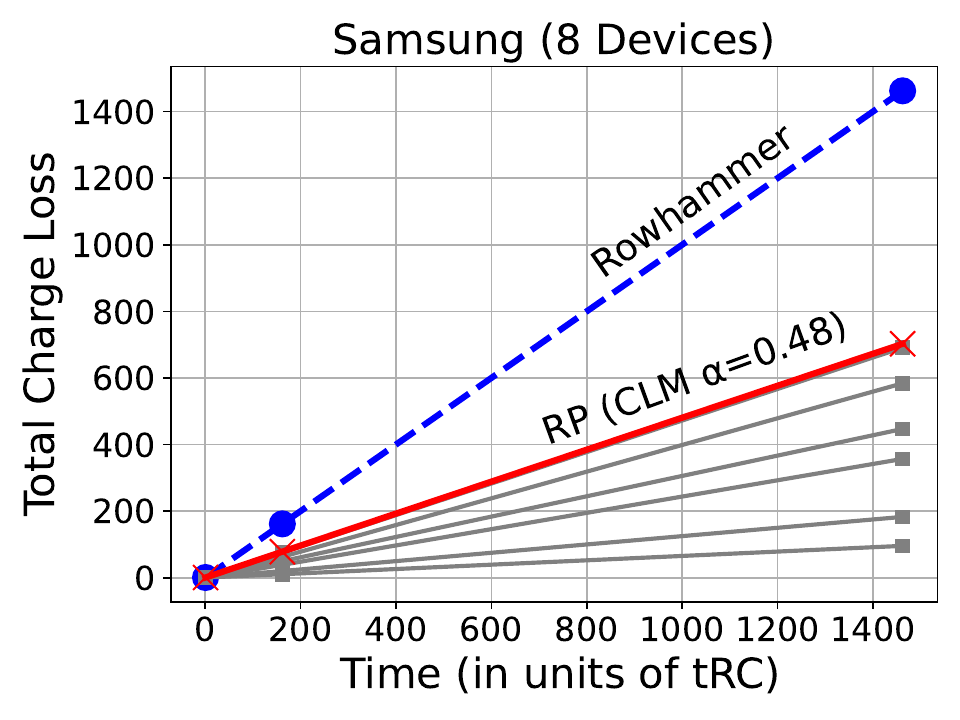}
\includegraphics[width=2in]{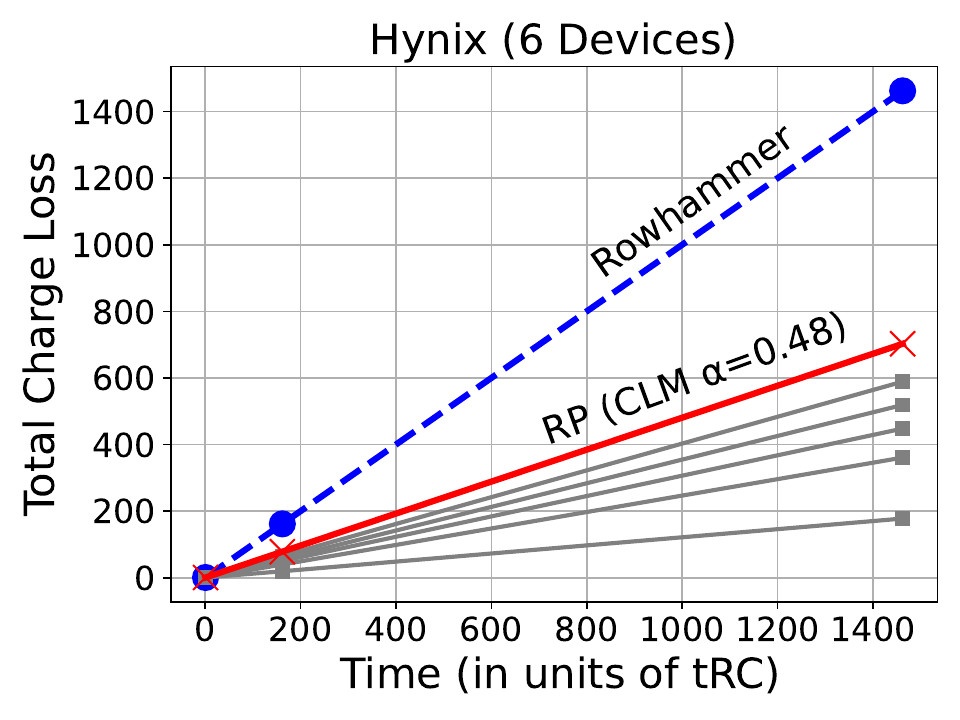}
\includegraphics[width=2in]{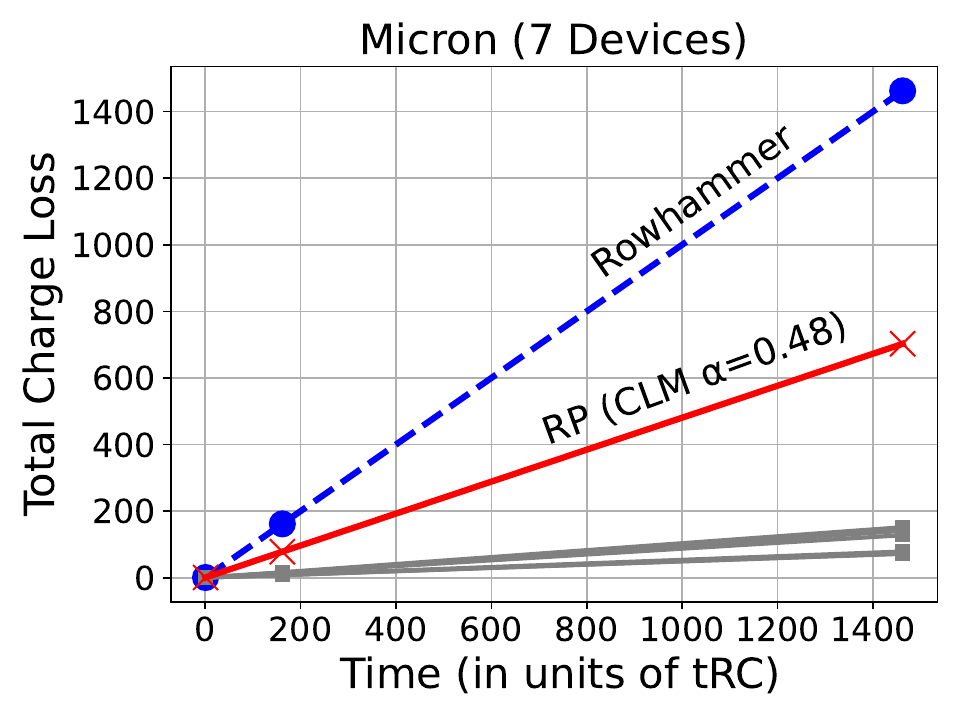}
    \vspace{-0.05in}
    \caption{Total Charge Loss (TCL) for long-duration RP attacks that last for 1 tREFI (162 tRC in DDR4) and 9 tREFI (1462 tRC in DDR4). For our CLM model for RP, we use alpha=0.48 as it covers all the devices across the three vendors (experimental data is reproduced from Appendix B of Luo et al.~\cite{rowpress})}
    \label{fig:tcl}
   \vspace{-0.15 in}
\end{figure*}

\begin{figure}[!htb]
    \centering
    \vspace{-0.05 in}
\includegraphics[width=2.75 in]{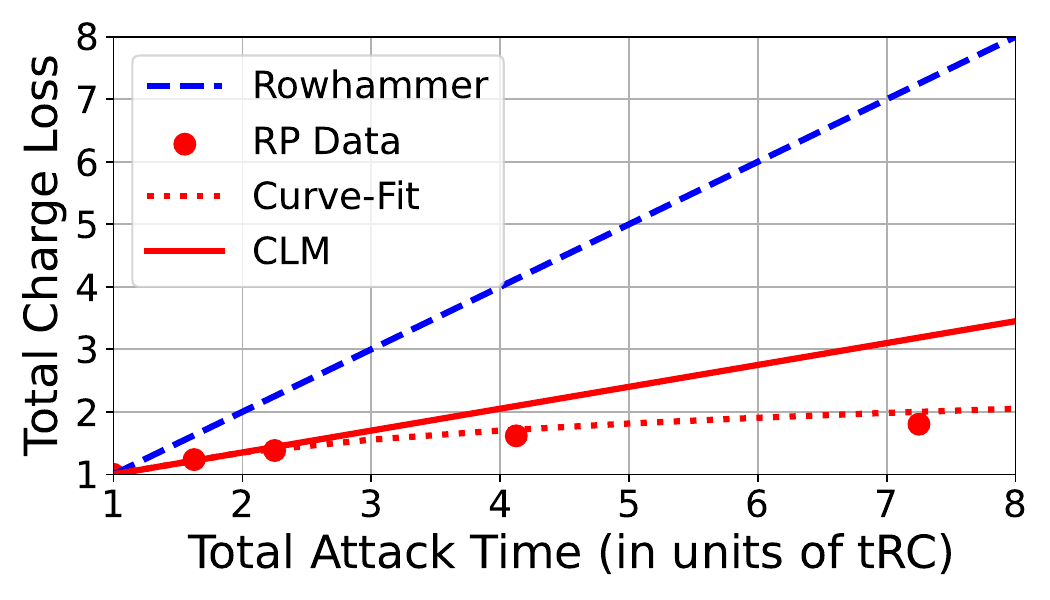}
\vspace{-0.1 in}
    \caption{Relative Charge-Loss Model for Row-Press} 
    \vspace{-0.1 in}
    \label{fig:rpmodel}
\end{figure}

If we have the data for T* available, we can deduce the relative charge leakage incurred by a single round of an RP attack (for a given tON time) compared to a single round of RH attack. For example, if the RP attack causes T* to be half of TRH, then each round of RP attack must leak 2x the charge as a single round of RH attack. We use this insight, to estimate the charge-leakage versus the attack time for an RP attack (note that the total time for an RP attack is tON+tPRE, as the attack eventually ends with a precharge). Figure~\ref{fig:rpmodel} shows the Total-Charge Loss for RP attack and compares it with  RH, as the attack time is increased from 1 tRC to 8 tRC.  RH is a linear attack (K units of charge-loss in K units of time).  The {\em red dots} represent the charge-loss derived from the data of Luo et al.~\cite{rowpress} (data is a reorganized version of Figure~\ref{fig:tonTRH}). 


\subsection{Conservative Linear Model (CLM)}

We could do a curve-fit on the experimental data (shown as dotted red line in Figure~\ref{fig:rpmodel}), however, we have two key requirements: (1) the function must be simple, as we want to use it for implementation in hardware, possibly inside the DRAM chip (2) the function must never underestimate the TCL observed in the chips, as underestimation can cause reliability and security failures if the actual loss is greater than the predicted loss. Based on these two constraints, we develop a {\em Conservative Linear-Model (CLM)}, which provides a linear relationship, albeit a conservative one.  Rather than looking for the best fit, and have an error in both directions, CLM produces a line, such that no observed data-point is above the line. The general form of CLM is given by Equation~\ref{eq:clm1}.

\begin{equation}
    TCL_{ON} = 1 + \alpha * ( \frac{tON-tRAS}{tRC} ) \label{eq:clm1} 
    \vspace{0.05 in}
\end{equation}

Where,  {\em $\alpha$ is the relative charge leakage per tRC for RP} ($\alpha$ of 1 gives RH).  For the data from Luo et al. (as shown in Figure~\ref{fig:rpmodel}), $\alpha$ equals 0.35, so Equation~\ref{eq:clm1} becomes Equation~\ref{eq:clm2}.

\begin{equation}
    TCL_{RPA}  = 1 + 0.35 * ( \frac{tON-tRAS}{tRC} )
    \vspace{0.05 in}
    \label{eq:clm2} 
\end{equation}

Note that RP attack degenerates into a RH attack if tON is equal to tRAS.  Thus, Equation~\ref{eq:clm1} represents a generalized equation that incorporates both RH and RP for any pattern.

\ignore{
performed a curve-fit on the characterization data, and the log-function provided the best fit (R-square of 0.97). For this curve, the TCL (shown with dotted-red line in Figure~\ref{fig:rpmodel}) is given by Equation X.

\begin{equation}
    TCL_{RPA} = 1 + 0.35 *Log2( \frac{tON-tRAS}{tRC} ) 
    \vspace{0.05 in}
\end{equation}
}

\subsection{Row-Press at Large Time Scale}

The experimental data shown in Figure~\ref{fig:rpmodel} is for a small-duration (sub-microsecond) RP attack. However, RP attacks can also be long-duration, lasting up-to one tREFI without refresh postponement and up-to 5x-9x times tREFI (DDR5-DD4) with refresh postponement.  Appendix-B of Luo et al.~\cite{rowpress} report characterization data for devices from all three memory vendors for the long-duration RP attacks, specifically  1 tREFI (162 tRC in DDR4) and 9 tREFI (1462 tRC in DDR4). Figure~\ref{fig:tcl} shows the Total Charge Loss (TCL) for those devices, as time is normalized in terms of tRC.  For comparison, the TCL of Rowhammer is also shown, if performed for an identical duration.  We also show our CLM model for RP, and we set $\alpha=0.48$, as it covers all the characterized devices.  Thus, we can use the model of Equations~\ref{eq:clm1} for modeling both short-duration and long-duration RP attacks.

\subsection{Key Observations:}

Our model allows us to estimate the combined effect (total charge loss) of an arbitrary pattern that interleaves RH and RP, where RP can have any length (limited by DDR specs). The key takeaways from our model are as follows:

 1. Row-Press is a much slower attack than Rowhammer. Even with $\alpha$ of 0.48, RP causes less than half the {\em damage} (charge-loss) per unit-time as a standalone RH attack. 

 2. Any time spent on RP is the time the attacker cannot perform RH. Therefore, solely doing RH is the fastest way to reach critical charge-loss, limited only by the RH mitigation.

 3. A secure RP solution must reduce dependency on $\alpha$ as $\alpha$ may vary across chips, or select the value of $\alpha$ conservatively, such that it is guaranteed to work across all the chips.

 4. As the leakage of RH is due to activity (row activation) and RP is due to idling, it is unlikely $\alpha$ would exceed 1. So, using $\alpha$ of 1  avoids the reliance on per-device behavior.

\ignore{

Overview ... explain with example

Impact limited to one tRC

Pattern

Impact

}

\newpage
\section{ImPress-N: The Naive Version}

We propose two variants of ImPress.  The first version is {\em ImPress-N}, the {\em naive} version, which is designed to handle only integer-values of charge-loss. The goal of ImPress-N is to understand the impact of reduced precision on the effectiveness of ImPress. ImPress-N divides the time into windows of tRC, and if a row is open for the entire window then it treats it as equivalent to an activation for the purpose of RH mitigation. Thus, ImPress-N limits the impact of any unmitigated Row-Press to at-most one tRC window. In this section, we provide the design and analysis of ImPress-N and bound the worst-case impact of the unmitigated Row-Press.

\subsection{ImPress-N: Design and Operation}

The key insight in ImPress-N is that secure RH mitigations are designed to tolerate the worst-case RH pattern, which causes an activation in each time window of tRC.  With Row-Press, if a row is kept open for a long time, then by design, such a pattern will not cause as many activations as the worst-case. If we convert the RP activity, into RH activity, then we can use the existing RH framework to mitigate RP. 

\begin{figure}[!htb]
    \centering
\includegraphics[width=3.4 in]{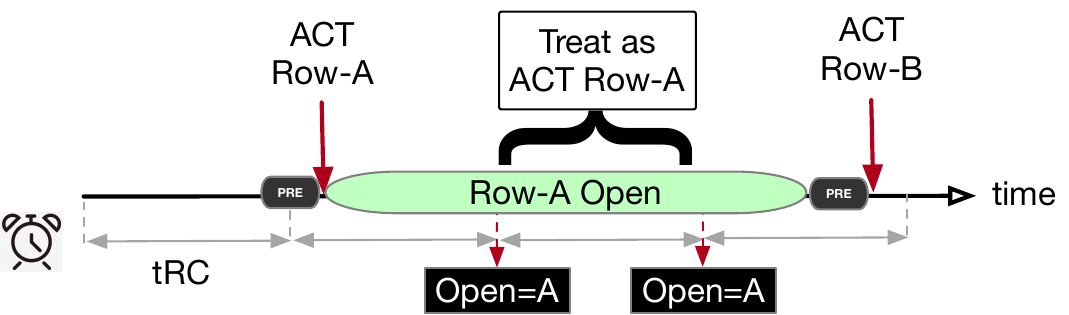}
    \caption{Design and Operation of ImPress-N. A row open for tRC is treated as equivalent to causing an activation within that window.} 
    \label{fig:design1}
\end{figure}

Figure~\ref{fig:design1} shows the overview and design of ImPress-N.  ImPress-N divides time into windows of tRC.  If a row activation occurs within the window, then that row participates in the RH mitigation. This is the case for Row-A in the second window and Row-B in the fourth window.  Furthermore, if a row is kept open for the entire tRC window, then it is treated as equivalent to causing a row activation for that open row, and that open row again participates in  RH mitigation. For example,  Row-A, which is open for tRC during the third window is treated as causing an activation on Row-A for the purpose of RH mitigation.

To implement ImPress-N, the system requires only two counters.  First, a {\em Timer} register that identify the ending time of each window. Second, an {\em Open-Row Address (ORA)} register that stores the row-address of the open row.  ORA is filled at the end of each window.  If the address to store in ORA is the same as the address present in ORA, it indicates that the row was open for the entire window, and participates in the RH tracking mechanism, similar to causing an activation.

ImPress-N is simple to incorporate in current RH-mitigation solutions, as it converts RP activity into a series of ACTs, which are already handled by RH-mitigation, so the underlying tracker design does not need to be changed. The total storage for implementing \textcolor{black}{ImPress-N is 1-byte for Timer, and 3-bytes for ORA, for a total of 4 bytes per bank (32 bytes per chip).}

\subsection{Bounding the Impact of Unmitigated Row-Press}

ImPress-N converts an RP pattern that keeps a row open over multiple tRC windows into an equivalent number of ACTs (one per tRC).  However, as it operates on integer values, it does not mitigate RP that occurs at the granularity of less than tRC. An attack can exploit this to reduce the threshold.

\begin{figure}[!htb]
    \centering
\includegraphics[width=3.4 in]{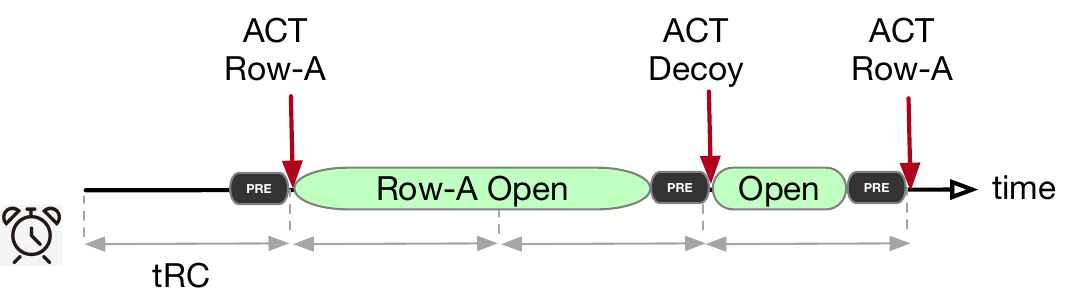}
    \caption{The pattern for exploiting the unmitigated Row-Press of ImPress-N -- an attacker can keep the row open for tRAS+tRC and evade RP mitigation.} 
    \label{fig:trhbound}
\end{figure}

Figure~\ref{fig:trhbound} shows the worst-case pattern for ImPress-N. The attacker is focused on causing undetected RP on Row-A.  The pattern causes an activation for Row-A at a time within the precharge-time (PRE) of the ending of the current window.  As Row-A is still not yet opened, it will not be stored in the ORA.  The pattern keeps Row-A open for a time equal to tRC+tRAS.  As Row-A is open at the end of the current tRC window, the address of Row-A is stored in ORA.  During the subsequent window, at a point slightly before the precharge time from the ending of the window, an ACT is sent for a decoy row, which causes precharge and closes Row-A.  Thus at the end of the window, ORA gets an invalid row. The pattern is repeated.

For each round of the pattern, the RH mitigation will see only a single ACT for Row-A, and thus treat this as a RH attack, causing a charge-loss of 1 per round for Row-A.  As the tON time for Row-A is  (tRC+tRAS), we can use Equation~\ref{eq:clm1} to quantify the charge loss per round as $(1+\alpha)$. Thus the {\em Effective Threshold  $(T^*)$} with ImPress-N is given by Equation~\ref{eq:impressd}.

\begin{equation}
    T^* = \frac{TRH}{(1+\alpha)}
    \label{eq:impressd}
\end{equation}

The impact on the threshold depends on $\alpha$.  The value of alpha from experimental data (tON $\le$ 2tRC) reported by Luo et al. is 0.35. So, T* is equal to TRH/1.35 or 0.74$\times$TRH. 
 If we want device independence, then $\alpha$=1 and T* equals TRH/2. 
 

\subsection{Protecting RH Trackers with ImPress-N}

Appendix-A describes how ImPress-N can be applied to our four tracker designs: PARA, Graphene, Mithril, and MINT.  For PARA and Graphene, we observe that both ExPress and ImPress-N have similar performance overheads as they both need to be operated at a reduced threshold (e.g. 2x lower). For Mithril and MINT, ImPress-N can make Row-Press mitigation viable at small performance overheads.  

\vspace{-0.05 in}
\begin{tcolorbox}
\noindent{\bf Key Takeaway:} For MC-based trackers, ImPress-N has a similar impact as ExPress on threshold, performance, and storage. However, as ImPress-N does not limit tON, it can also be used with in-DRAM trackers, thus representing the first solution to protect such trackers from RP attacks.  
\end{tcolorbox}
\vspace{-0.05 in}

\clearpage
\section{ImPress-P: The Precise Version}

While ImPress-N is a simple design (no changes to the trackers, except for the number of entries), it can still incur performance overheads due to the lowering of the effective threshold resulting from unmitigated Row-Press that occurs at sub-tRC granularity.  Furthermore, the impact of ImPress-N on the threshold depends on the value of $\alpha$, and we want a solution that naturally offers protection of $\alpha$=1 without any of the associated overheads. Our next design, {\em Impress-P (Precise)}, overcomes both shortcomings of ImPress-N. The key idea in ImPress-P is to measure the tON time of a row, and use it to determine the {\em Equivalent Number of Activations (EACT)} between the time the row is opened and it completes precharge. ImPress-P ensures that there is no lowering of the threshold due to mitigating Row-Press. In this section, we provide the design and analysis of ImPress-P and study the impact of applying Impress-P to different trackers.

\subsection{ImPress-P: Design and Operation}

The key insight in ImPress-P is that secure RH mitigations are designed to tolerate the rate of damage that occurs under the RH pattern.  So, we can treat every time unit in terms of tRC (integer or fractional) as equivalent to that amount of ACTs (integer or fractional).  This allows us to convert any amount of RP activity precisely into equivalent RH activity, and use the existing RH framework to accurately mitigate RP without impacting the threshold.

\begin{figure}[!htb]
    \centering
\includegraphics[width=3.25 in]{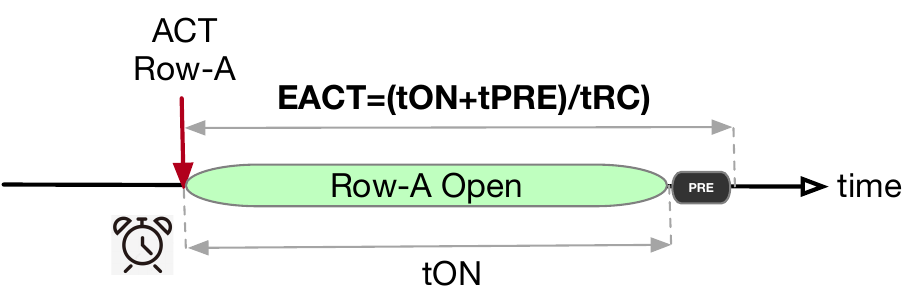}
    \caption{Design and Operation of ImPress-P. ImPress-P measures the time the row is open and converts it into an equivalent number of ACT (EACT).} 
    \label{fig:design2}
\end{figure}

Figure~\ref{fig:design1} shows the design of ImPress-P.  ImPress-P only requires a timer to measure the amount of time the row is open (tON). The timer starts when the row is opened and stops when the row is closed. The total duration for access must also include the time required for precharge, so the total time equals (tON+tPRE).  We divide the total time by tRC to get the {\em Equivalent Number of ACTs} {\em (EACTs)}.  For example, if tON is equal to tRAS, this is the same as RH attack, and EACT is equal to 1.  If tON is equal to tRAS+tRC, the access lasts for two tRC and we would get EACT=2.  EACT is guaranteed to be at-least 1 but it can be a fractional value (e.g.  if tON=tRAS+tRC/2, EACT=1.5).  Thus, the RH-mitigation algorithms must be able to handle non-integer number of ACT.

For counter-based algorithms, we modify the counters to support fractional value, and instead of incrementing by 1, we increment the counter by EACT.  For probabilistic solutions, we modify the selection probability from $p$ to $p * EACT$. Thus, ImPress-P is applicable to both types of trackers. 

ImPress-P requires \rev{a single {\em Timer} (10-bits) per bank (32 per chip)}.  All DRAM activity occurs and is measured at the granularity of DRAM cycles. For our 2.66GHz DRAM, this means tRC (48ns) is equal to 128 cycles, thus the division by tRC can be implemented by shifting right by 7 bits.

\subsection{Impact of Counter Precision on Effective Threshold}

The fractional part of EACT is 7-bits (due to division by tRC). For the counter-based tracking algorithms, this means the counter must also be extended by 7 bits to precisely incorporate the fractional values of EACT. A design may choose to modify the counter-based tracker with fewer bits to store the fractional value (to save on storage) at the expense of some amount of error in tracking, which leads to an equivalent reduction in the effective threshold (T*).

\begin{figure}[!htb]
    \centering
\includegraphics[width=3.5 in]{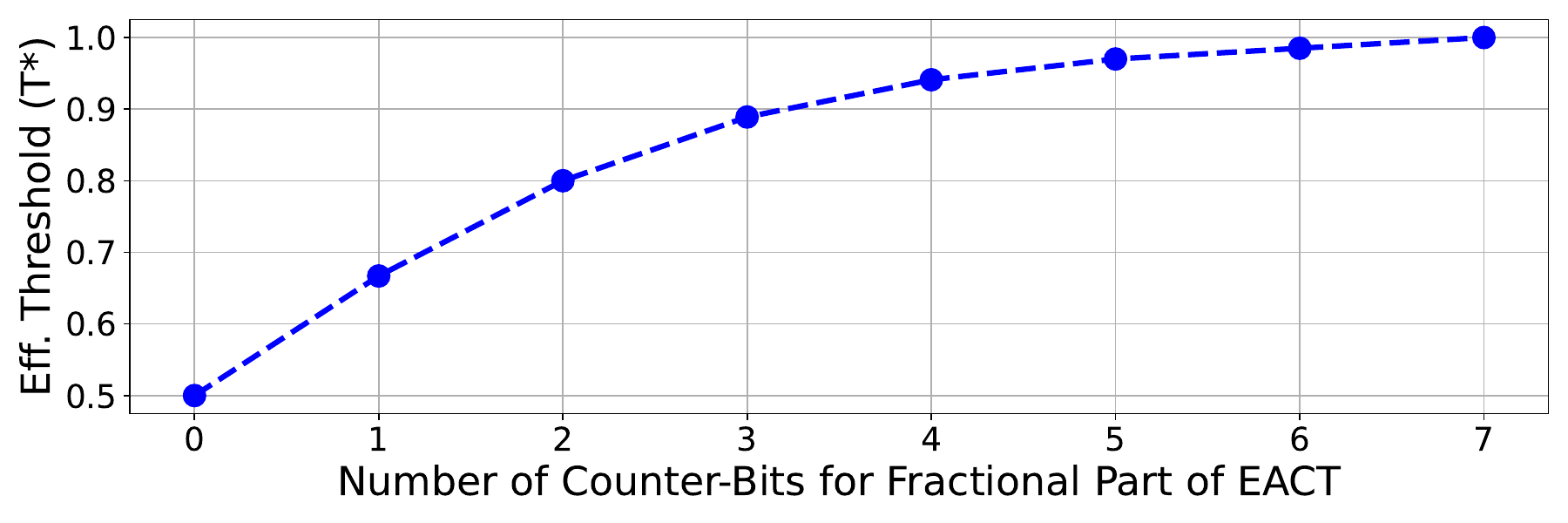}
\vspace{-0.2 in}
    \caption{Impact of number of counter-bits for storing the fractional part on the effective threshold of ImPress-P (value is normalized to TRH). } 
    \label{fig:precision}
\end{figure}

Figure~\ref{fig:precision} shows the effective threshold (T*) of ImPress-P as the number of counter-bits used for storing the fractional part is varied from 0 to 7.  With 7-bits we track accurately, so T* is equal to TRH (no reduction in threshold). With fewer than 7-bits, say $b$ bits, we get a precision equal to $\frac{1}{2^b}$, so the loss in accuracy is also equal to $\frac{1}{2^b}$. Thus, with 6-bits, the relative T* reduces to 0.985, with 5-bits to 0.97, with 4-bits to 0.94. Finally, if we have 0-bits for the fractional part, ImPress-P degenerates to ImPress-N, and has T* of 0.5 times TRH.

Our default implementation of ImPress-P uses 7-bits for storing the fractional part.  Thus, ImPress-P maintains the {\em same} TRH with Row-Press protection as compared to a system without any Row-Press protection. Furthermore, ImPress-P avoids any dependency on $\alpha$ (it is implicitly designed for $\alpha$ of 1).  Thus, while implementing and comparing designs with ImPress-P, we will use $\alpha$=1.

\subsection{Protecting RH Trackers with ImPress-P}

Unlike ExPress, ImPress-P does not place any limit on tON.  Thus, it does not impact performance due to early closure of an open row. Furthermore, as ImPress-P does not affect the threshold, it also does not incur any additional mitigations due to activations compared to an idealized baseline that does not have Row-Press.  However, ImPress-P can still incur additional mitigations due to a row being kept open for a long time.

We analyze ImPress-P, ImPress-N and ExPress for our trackers. We implement ExPress with tMRO of tRAS+tRC.  As ExPress is incompatible with in-DRAM tracker designs (Mithril and MINT), we compare ImPress-P with only ImPress-N for these two designs. We describe the changes required in the tracking algorithms to support ImPress-P.

\begin{figure*}[!tbh]
    \centering
     \vspace{-0.1 in}
    \includegraphics[width=0.89\textwidth]{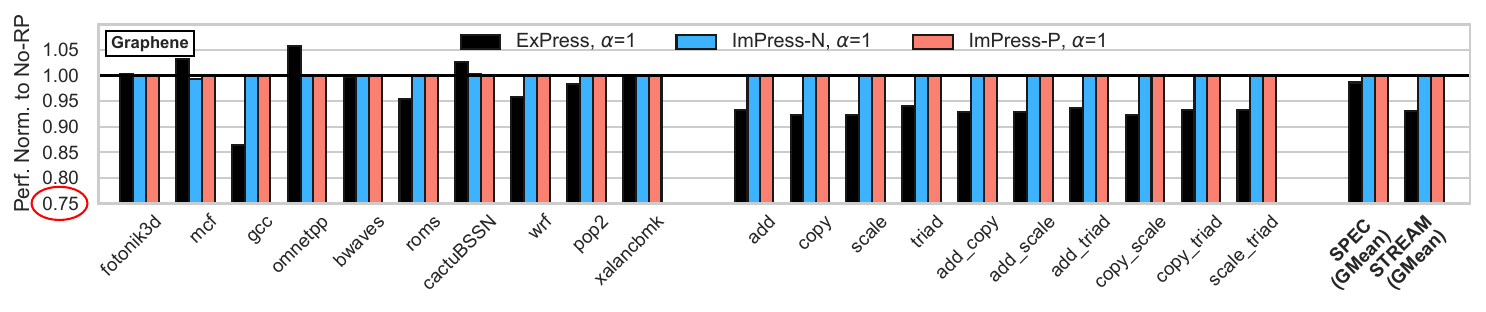}
    \vspace{-0.1 in}
    \includegraphics[width=0.89\textwidth]{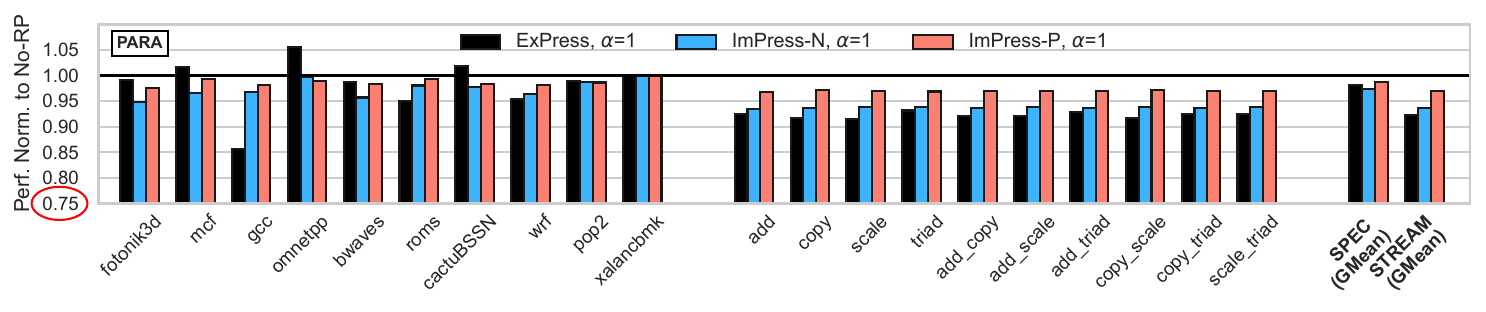}
    \vspace{-0.08 in}
    \includegraphics[width=0.89\textwidth]{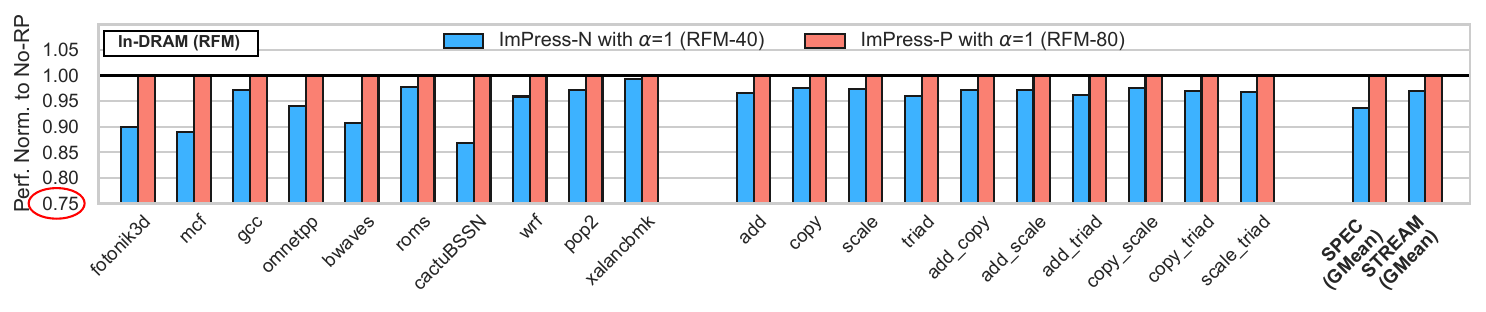}
    \vspace{-0.1in}
    \caption{Performance of (a, top) Graphene and (b, mid) PARA for ExPress, ImPress-N, and ImPress-P  (c, bottom) Performance of in-DRAM  (MINT) for ImPress-N and ImPress-P (ExPress is not shown as it is not applicable to in-DRAM trackers). Note: All performance is normalized to No-RP.}
    \label{fig:perf-impressp}
    \vspace{-0.15in}

\end{figure*}
\newpage


\noindent{\bf Impact on Graphene:} For TRH of 4K,  Graphene requires 448 entries per bank. Both ExPress and ImPress-N (\alfa of 1) doubles it to 896 per bank.  With ImPress-P, the number entries remains unchanged at 448. However, each entry now requires 7-bits of extra storage to store fractional value of EACT, hence ImPress-P incurs 25\% storage overhead (each entry is 28-bits). Thus, the total storage required for ImPress-P is only 1.25x of No-RP, whereas it was 2x for both ImPress-N and ExPress.


Figure~\ref{fig:perf-impressp} shows the performance of Graphene with ExPress  ImPress-N, and ImPress-P, normalized to a baseline that does not suffer from RowPress. As ImPress-P does not affect the threshold or restrict tON, it incurs a negligible  overhead.

\vspace{0.05 in}

\noindent{\bf Impact on PARA:} Conventionally, PARA uses a constant probability {\em p} for all activations. For TRH of 4K, p=1/184, and for ImPress-N and ExPress p=1/92.  ImPress-P changes PARA to use a variable value for $p$ for each activation, depending on the tON time. For each activation, PARA uses $\hat{p}=p*EACT$.



Figure~\ref{fig:perf-impressp} shows the performance of PARA with ExPress, ImPress-N, and ImPress-P, normalized to a baseline without Row-Press. ImPress-P has significantly reduced performance overheads (especially for Stream) compared to ExPress.

\ignore{
\vspace{0.05 in}

As ExPress is not compatible with in-DRAM trackers, we compare ImPress-P only with ImPress-N, for evaluating Mithril and MINT. For in-DRAM trackers, to implement ImPress-P, each bank would provision the {\em Timer} register and perform the computation for EACT (the division by tRC is equivalent to a shift by 7 bits, so it is easily implementable). 
}


\vspace{0.05 in}
\noindent{\bf Impact on Mithril:} For TRH of 4K, and a default RFMTH of 80, Mithril requires 383 entries.  This increases to 1545 entries (4x) with ExPress and ImPress-N ($\alpha$=1).  With ImPress-P, the number of tracking entries remains unchanged at 383. However, each entry must now be provisioned with 7 more bits to track the fractional values, resulting in 25\% storage overheads, much less than the 4x overhead required for ExPress and ImPress-N. The performance overheads of Mithril, due to RFM commands, remain the same as No-RP baseline. 


\vspace{0.05 in}
\noindent{\bf Impact on MINT:} MINT contains three registers: SAN (Selected Activation Number), CAN (Current Activation Number), and SAR (Selected Address Register).  Both SAN and SAR remain unchanged.  We modify CAN to have 7 more bits corresponding to the fractional value of EACT.  For each activation, we increase CAN by the value of EACT. Thus, each activation gets a selection probability in proportion to the EACT. If CAN crosses SAN, the row-address is stored in SAR.  At RFM, the row-address in SAR (if valid) is mitigated and a new value for SAN is selected. ImPress-P increases the storage overhead of MINT from 4 bytes to 5 bytes. 
  With ImPress-N, the threshold increases from 1.6K to 3.1K, whereas with ImPress-P, it remains unchanged at 1.6K. Figure~\ref{fig:perf-impressp}(c) shows the performance of No-RP, ImPress-P and ImPress-N. ImPress-P has an identical performance to No-RP.


\subsection{Summary of Comparisons}

Table~\ref{table:compare} compares ExPress, ImPress-N, and ImPress-P. The shortcomings are highlighted in {\bf{bold}}.  ImPress-P requires minor changes (to include EACT) and provides near-ideal performance.  Therefore, we will assume that by default ImPress is implemented only as ImPress-P (ImPress-N was an intermediate step to emphasize the importance of precision).


\begin {table}[htb]
\begin{scriptsize}
\begin{center} 
\vspace{-0.1in}
\caption{Comparisons of ExPress, ImPress-N, and ImPress-P}
\vspace{-0.1in}
\begin{tabular}{cccc}
\hline

 Property	&	ExPress	&	ImPress-N	&	ImPress-P	\\ \hline
Puts Limit on tON	     &	\redd{Yes}	&	No	&	No	\\
Affects Threshold (T*)	 &	\redd{Yes (up to 2x)}	&	\redd{Yes (up to 2x)}	&	No (1x)	\\
Performance  Overheads   &	\redd{High}	&	Medium	&	Low	\\
More Tracking Entries	 &	\redd{Yes (up to 2x)}	&	\redd{Yes (up to 2x)}	&	No (1x)	\\
Wider Tracking Entries	 &	No	&	No	&	\redd{Yes (Minor)}	\\
In-DRAM Trackers	    &	\redd{Incompatible}	&	Compatible	&	Compatible	\\
Device Dependency	    &	\redd{Yes (alpha)}	&	\redd{Yes (alpha)}	&	No	\\ \hline

\end{tabular}
\label{table:compare}
\vspace{-0.25 in}
\end{center}
\end{scriptsize}
\end{table}

\subsection{Activation and Energy Overheads}

\begin{figure}[!htb]
    \centering
\vspace{-0.1in}
\includegraphics[width=3.5 in]{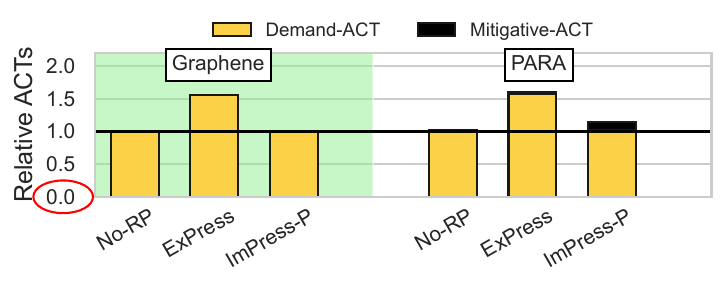}
\vspace{-0.3 in}
    \caption{Relative activation overhead of Graphene and PARA (No-RP, Express, and Impress-P), broken down into demand activations and mitigative activations (all normalized to activations in the unprotected baseline).} 
    \label{fig:act_overhead}
\end{figure}

Tolerating Row-Press can cause extra activations due to row closure (ExPress) or additional mitigations.
\cref{fig:act_overhead} shows the activations, averaged over all workloads, relative to an unprotected baseline. 
 Graphene without RP protection (No-RP) causes less than 1\% extra activations.
With ExPress, mitigative activations remain low, but demand activations increase by 56\% in Graphene (57\% for PARA). 
Graphene with ImPress-P incurs no additional activation overhead.

For PARA, the extra demand activations with ImPress-P are negligible at 2\% on average, however, the mitigative activations increase by 12\%.
Overall, ImPress-P has significantly lower activation overhead compared to ExPress, reducing it from 56\% to 1\% for Graphene, and 61\% to 14\% for PARA.

\noindent{\bf Energy Overheads:} On average, activations account for 11\% of the baseline DRAM energy. ExPress increases DRAM energy by 6\% for Graphene (7\% for PARA), while for Impress-P, the increase in energy is 1\% for Graphene (2\% for PARA).

\subsection{Scalability to Lower Rowhammer Threshold}
\label{sec:TRH}

\begin{figure}[!htb]
    \centering
\vspace{-0.1 in}
\includegraphics[width=3.5 in]{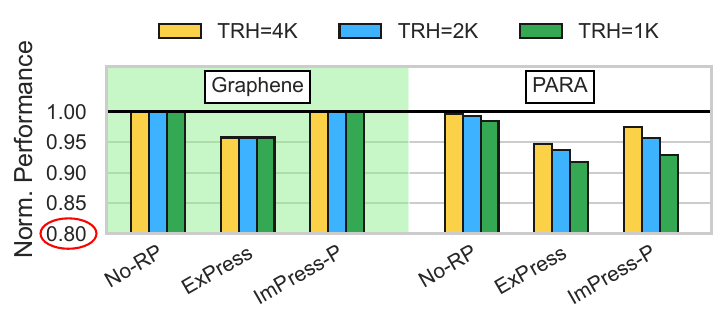}
\vspace{-0.3 in}
    \caption{Performance of Row-Press mitigation with Graphene and PARA, normalized to an unprotected baseline. [Note: All values are Geo-Mean.]} 
    \label{fig:thres_sens}
\vspace{-0.15 in}
\end{figure}

\cref{fig:thres_sens} shows the performance of Graphene and PARA normalized to an unprotected baseline as TRH varies from 1K to 4K.
At TRH of 1K.
Graphene incurs no slowdown for No-RP and ImPress-P, while ExPress has 4.4\% slowdown.
PARA incurs 1.5\% slowdown for No-RP and ExPress increases the slowdown to 8.9\%. ImPress-P reduces it to 7.7\%. The storage overheads of Graphene and performance overheads of PARA make them impractical for low TRH. For low TRH, some companies~\cite{bennett2021panopticon, HynixRH} and JEDEC~\cite{prac} (announced one day before MICRO submission deadline) are adopting {\em Per-Row Activation Counting (PRAC)} where the DRAM array stores a counter for each row (8KB). ImPress can be used with PRAC by having 7-bits of the counter for storing the fractional EACT.

\ignore{

RH Tracking

RH Mitigating Actions

Modifying DRAM to Improve:  REGA, HiRA

Error Correction for RH: SafeGuard and CSI-Rowhammer

General works in DRAM Reliability 

ArchShield, XED, ... On-Die ECC paper, Mattan paper's

}

\section{Related Work}

\ignore{
In this paper, we focus on Data-Disturbance Errors (DDE) in DRAM.  While Rowhammer has been well-studied, Row-Press is a recent discovery, with the seminal paper from Luo et al.~\cite{rowpress} being the key work.  We discuss that work extensively and compare our proposal to the solution presented in their paper.  Now, we describe other related work, especially those related to tolerating the problem of Rowhammer. 
}

\color{black}

To the best of our knowledge, ImPress-P represents the first tracker implementation that can securely tolerate both Rowhammer and Row-Press. ImPress exploits the observation that the row-open time can be converted into equivalent activity for Rowhammer. Prior works have made similar observations. For example, ProTRR~\cite{ProTRR} was the first to suggests to ``increase the counter for victims of the (aggressor) row that remains active". However, it does not provide methodology to convert the row-open time into equivalent RH (notably, ProTRR appeared one year before Row-Press was publicly known and characterized, so the lack of such details is understandable).  Furthermore, ProTRR operates with integer-valued counters, and ImPress-N shows that such an integer-valued design has a significantly higher threshold.

While DSAC~\cite{DSAC} uses {\em time-weighted} counting, it suffers from three problems: (1) the weight is a logarithmic function of time, for example, for tON=256 tRC, the weight will be approximately 8, whereas, the Row-Press characterization~\cite{rowpress} shows that the weight should be about 0.48*256 = 122 (15x higher).  Thus, DSAC significantly underestimates the RP damage, (2) Row-Press is ignored for the row getting installed in the tracker, as it always uses a weight of 1, (3) DSAC  uses integer counter values and would suffer from the same problem as ImPress-N, even if the weights were accurate. We note that DSAC  can be broken with Blacksmith~\cite{proteas}, so assessing the security of DSAC against Row-Press is impractical.

\color{black}


 Several studies~\cite{kim2014architectural}~\cite{kim2014flipping}~\cite{MRLOC}~\cite{PROHIT}~\cite{kim2014architectural}~\cite{cbT}~\cite{lee2019twice},~\cite{park2020graphene}~\cite{qureshi2022hydra} have investigated efficient trackers to identify aggressor rows. Our design can work with any of these trackers. We do not consider In-DRAM designs of TRR~\cite{frigo2020trrespass}, DSAC~\cite{DSAC}, and PAT~\cite{HynixRH} as these can be broken with simple patterns~\cite{frigo2020trrespass}\cite{jattke2021blacksmith}. Our work is applicable to secure in-DRAM trackers, such as Mithril~\cite{kim2022mithril}, MINT~\cite{kim2022mithril}, ProTRR~\cite{ProTRR}, and PRHT~\cite{HynixRH}.

Prior works have looked at alternative mitigation techniques, such as rate-limiting~\cite{yauglikcci2021blockhammer} or  Dynamic row-migration~\cite{saileshwar2022RRS}~\cite{AQUA}~\cite{SRS}~\cite{ShadowHPCA23}. Prior studies~\cite{cojocar2019eccploit,ali2022safeguard,csi,DSN23_PTGuard,twobirds} have also proposed to use ECC and detection codes to tolerate Rowhammer. All these works can reduce, but not eliminate, DDE errors.  REGA~\cite{REGA_SP23} and HiRA~\cite{HIRA} modify the DRAM module to support multiple concurrent mitigative activations.

\section{Conclusion}

The scaling of DRAM to single-digit nanometers results in new modalities of {\em Data-Disturbance Errors (DDE)}. While Rowhammer is well-known, recently, a new pattern, Row-Press, was discovered, which causes charge leakage by keeping the row open for a long time. RP reduces the number of activations required to induce a bit-flip. Prior work proposed to mitigate RP by limiting the maximum time a row can be kept open, however, that proposal incurs high overheads and is incompatible with in-DRAM tracking. We propose {\em Implicit Row-Press (ImPress) mitigation}, which converts RP activity into an equivalent amount of RH activity, and uses the RH framework to mitigate RP.  Our solution does not restrict tON, incurs low overheads, and is applicable to all trackers. 


\begin{figure*}[!tbh]
    \centering
\includegraphics[width=0.89\textwidth]{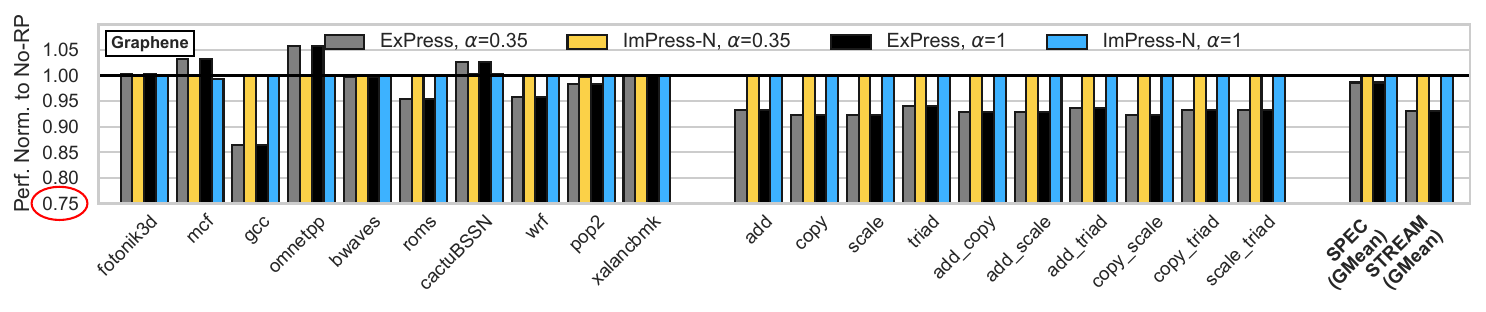}
    \vspace{-0.1in}
    \includegraphics[width=0.89\textwidth]{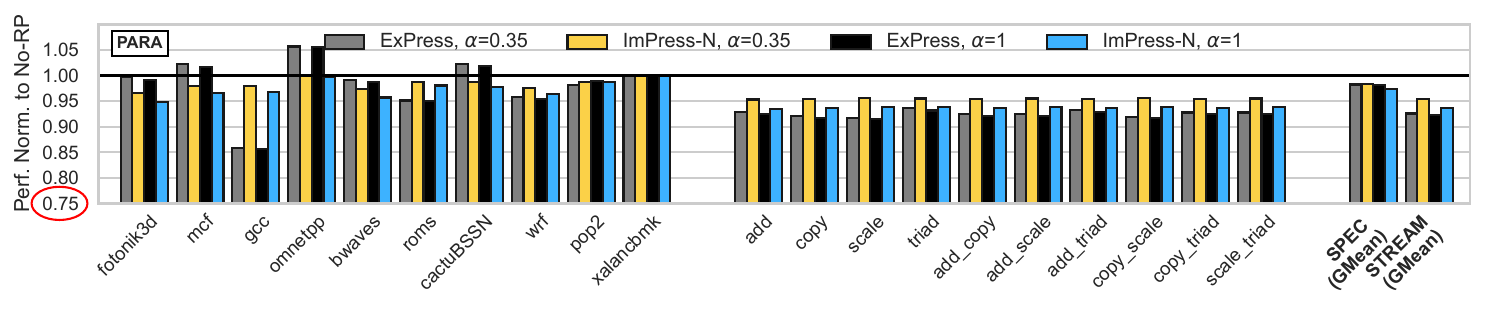}
    \vspace{-0.08in}
    \includegraphics[width=0.89\textwidth]{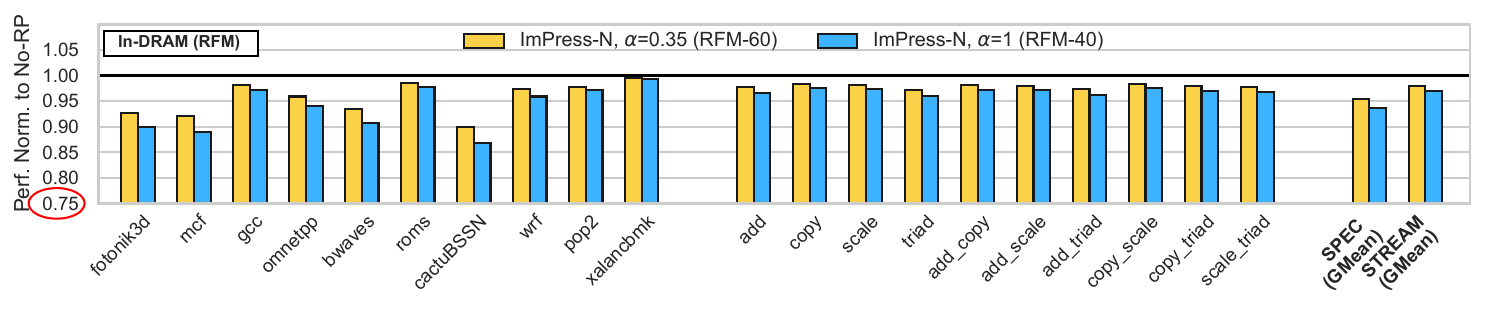}
    \vspace{-0.08in}\caption{Performance of (a, top) Graphene and (b, mid) PARA for ExPress and ImPress-N (both designs for $\alpha$ of 0.35 and 1) (c, bottom) Performance of in-DRAM  (MINT) for ImPress-N (ExPress is not shown as it is not applicable to in-DRAM trackers). Note: All performance is normalized to No-RP.}

    \label{fig:perf-impressd}

\end{figure*}

\newpage
\section*{Appendix-A: Performance Impact of ImPress-N}
Unlike ExPress~\cite{rowpress}, ImPress-N does not place any limit on tON, therefore it does not suffer from reduced row-buffer hits due to premature closing of an open row due to tMRO.  However, ImPress-N still incurs performance overheads from to the extra mitigations due to the reduction in threshold (T*) and from considering rows opened for tRC as an ACT. 

We analyze ImPress-N and ExPress for our four trackers.  To ensure that both schemes are targetted to the same T*, we evaluate ExPress with tMRO set to (tRAS+tRC).  


\vspace{0.05 in}
\noindent{\bf Impact on Graphene:} For TRH of 4K, Graphene uses an internal threshold of 1333 (mitigation is sent when counters reach the internal threshold), requiring  448 entries per bank (a total of 115KB SRAM per channel). To make Graphene Row-Press tolerant with ExPress or ImPress-N, the number of entries must be increased in direct proportion to (1+$\alpha$).  Thus, for \alfa of 0.35, Graphene requires 605 entries per bank (a total of 155KB SRAM per channel), and \alfa of 1, Graphene requires 896 entries per bank (a total of 230KB SRAM per channel).  Thus, both ExPress and ImPress-N require a total storage overhead of 1.35x-2x compared to the No-RP design.

Figure~\ref{fig:perf-impressd} shows the performance of Graphene with ExPress and ImPress-N, normalized to No-RP.  As Graphene is efficient in sending mitigative refreshes, the slowdown mainly comes from the reduction in row-buffer hits. For Stream workloads, ExPress incurs an average slowdown of 7.5\%, whereas ImPress-N incurs a negligible slowdown. For SPEC, both ExPress and ImPress-N have similar performance.

\vspace{0.05 in}

\noindent{\bf Impact on PARA:} For TRH of 4K, PARA requires {\em p} to be 1/184.  At \alfa of 0.35, $p$ increases by 1.35x to 1/136, for both ExPress and ImPress-N.  At \alfa of 1, $p$ increases to 1/92 for both ExPress and ImPress-N. Figure~\ref{fig:perf-impressd} shows the performance of PARA with ExPress and ImPress-N, normalized to No-RP.  On Stream workloads, ExPress incurs an average slowdown of 8\%  (at \alfa of 0.35) and 8.4\% (for \alfa of 1), whereas, ImPress-N incurs an average slowdown of 4.7\%  (at \alfa of 0.35) and 6.7\% (for \alfa of 1). Overall, ImPress-N performs better than ExPress.


As ExPress is incompatible with in-DRAM trackers,   we evaluate Mithril and MINT only with ImPress-N. 

\vspace{0.1 in}

\noindent{\bf Impact on Mithril:} We assume a default {RFM Threshold (RFMTH)} of 80. For such RFMTH, to handle a TRH of 4K, Mithril requires 383 entries.  To account for the unmitigated RP of ImPress-N, Mithril would need to target a revised threshold (T*) of either 2963 ($\alpha${}=0.35) or 2000 ($\alpha${}=1). Thus, the number of entries increases from 383 to 615 ($\alpha${}=0.35) or 1545 ($\alpha${}=1). 

We assume a system that already performs RFM at RFMTH of 80 (to tolerate Rowhammer). Therefore, Mithril and MINT do not incur any additional performance overheads.

\ignore{

\begin{figure}[!htb]
    \centering
    \includegraphics[width=0.9\columnwidth]{Graphs/raa_10M.pdf}
   \vspace{-0.1 in}
    \caption{\anishTODO{Performance of in-DRAM Trackers (Mitrhil and MINT) with RFM.}} 

    \label{fig:rfm}
\end{figure}
}

\vspace{0.1 in}

\noindent{\bf Impact on MINT:} For MINT, we use RFMTH of 80. Therefore, for No-RP, MINT can tolerate a TRH of 1.6K.  Due to the unmitigated Row-Press of ImPress-N, the tolerated threshold increases to 2.1K ($\alpha${}=0.35) and  3.1K ($\alpha${}=1). Alternatively, we could reduce RFMTH to 60 ($\alpha${}=0.35) or 40 ($\alpha${}=1) to retain the same tolerated TRH (of 1.6K).  Figure~\ref{fig:perf-impressd} shows the slowdown of RFM-60 and RFM-40 compared to RFM-80. The average slowdown is small, and ranges from 3\% to 5\%.

\clearpage

\section*{Appendix-B: Performance Impact  of Attacks}

We are interested in analyzing the performance implications of ImPress-P under attacks that combine both Rowhammer and Row-Press. We note that such patterns affect the performance only for memory-side mitigations (the performance of in-DRAM Rowhammer mitigations remains independent of the access patterns as mitigations are performed under REF).

\subsection{The Parameterized Attack Pattern for RH and RP}

\begin{figure}[!htb]
    \centering
    \vspace{-0.05 in}
\includegraphics[width= 3 in]{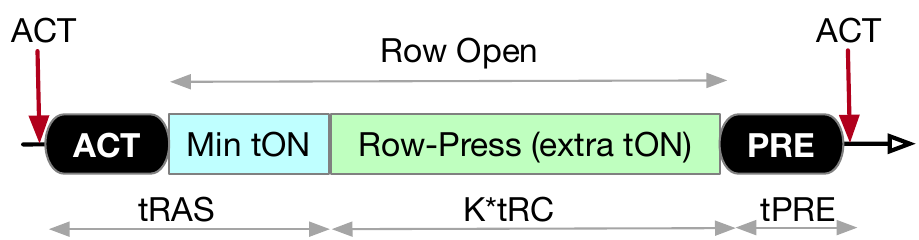}
    \caption{Attack Loop for combined Rowhammer and Row-Press pattern.}
    \vspace{-0.1 in}
    \label{fig:attackloop}
\end{figure}

We form a parameterized version of the pattern, as shown in Figure~\ref{fig:attackloop}. The pattern contains an activation that keeps the row open for tRAS.  Then, the row is kept open for an additional $K$ times tRC time-period, where K is the Row-Press parameter.  Finally, the row is closed incurring tPRE time.  Thus, the total time for one loop is (K+1)*tRC. If K=0 this pattern becomes Rowhammer.  If K=1, this pattern is a short-duration Row-Press.  If K=72, this pattern keeps the row open for a full tREFI. The pattern is repeated continuously and we are interested in the relative time taken to perform a large number of attack iterations (N).

\subsection{Analyzing the Performance Impact on Graphene}

\begin{figure}[!htb]
    \centering
    \vspace{-0.1 in}
\includegraphics[width= 3.6 in]
{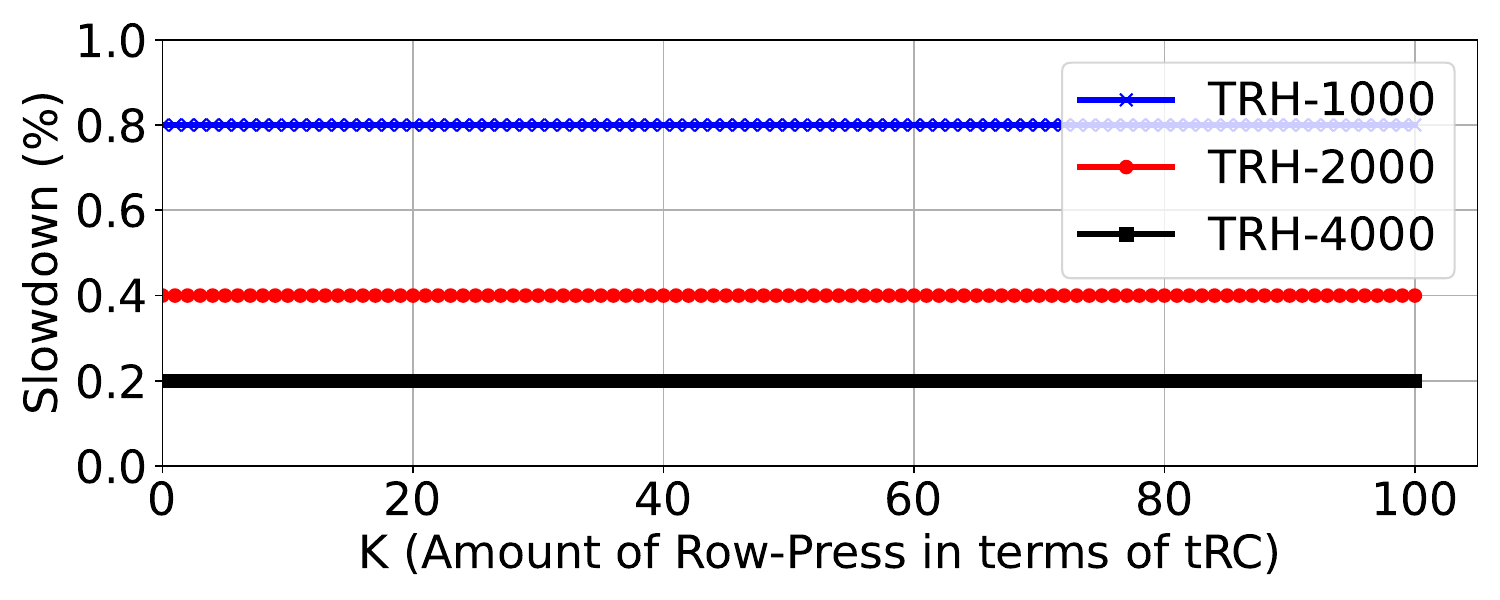}
\vspace{-0.25 in}
    \caption{Slowdown of ImPress-P with Graphene for the attack pattern.}
    \vspace{-0.05 in}
    \label{fig:grapheneattack}
\end{figure}

Let $T$ be the Rowhammer threshold.  Graphene performs mitigation when the counter reaches $T/2$ activations (due to periodic reset of the counter). For each mitigation, we need 4 activations (Blast Radius is 2, so 2 victim rows on each side of an aggressor row). Thus, the throughput loss under Rowhammer attack is  4/(T/2) or 8/T. The slowdown is 0.2\%/0.4\%/0.8\%  for T=4000/2000/1000, as shown in Figure~\ref{fig:grapheneattack}.

For analyzing Row-Press, we vary K. The total time for N iterations in the nonsecure baseline is N * (K + 1) * tRC. In each iteration of the loop, the counter of Graphene will increase by (K+1).  When it reaches T/2, Graphene will issue a mitigation (4 activations). Thus, the slowdown will be 4 activations per (T/2)/(K+1) iterations of the loop. To simplify our analysis and without loss of generality, consider the case where the attack loop is repeated N = (T/2)/(K+1) times. 

\begin{equation}
    t_{mitigation} = 4 \cdot tRC
\end{equation}
\begin{equation}
   t_{one-iter}  = (K+1) \cdot tRC   
\end{equation}
\begin{equation}   
    t_{N} = (K+1) \cdot tRC \cdot 
    \frac{(T/2)}{(K+1)} = (T/2)\cdot tRC  
\end{equation}
\begin{equation}   
    Slowdown = t_{mitigation}/t_{N} = \frac{4\cdot tRC}{(T/2)\cdot tRC} = 8/T
\end{equation}

Thus, the slowdown of Graphene remains 8/T, independent of ``K" (any amount of Row-Press).  This is expected as ImPress-P converts Row-Press into an equivalent amount of Rowhammer; therefore, the slowdown per unit time of attack remains the same, regardless of whether the pattern is Rowhammer or Row-Press. Figure~\ref{fig:grapheneattack} shows the slowdown of Graphene as the amount of Row-press is varied (for TRH of 1000/2000/4000). For each threshold, the slowdown remains independent of Row-Press.

\subsection{Analyzing the Performance Impact on PARA}

\begin{figure}[!htb]
    \centering
    \vspace{-0.05 in}
\includegraphics[width= 3.6 in]{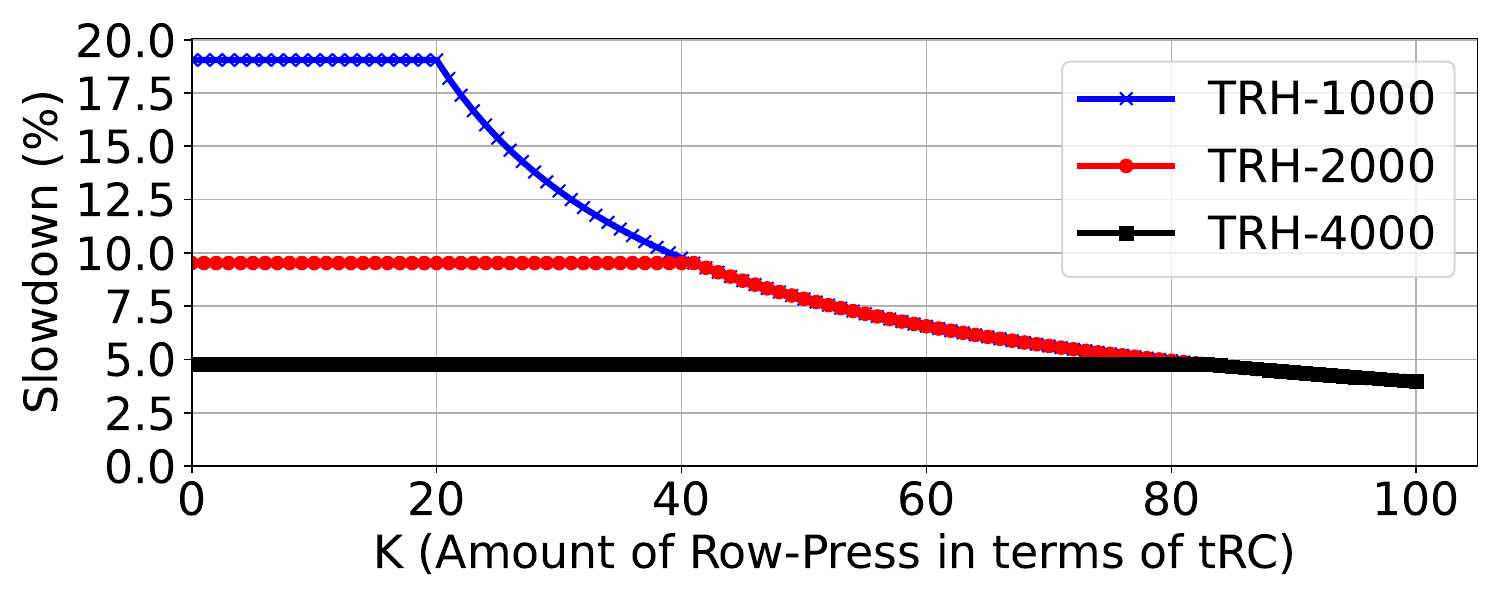}
\vspace{-0.25 in}
    \caption{Slowdown of ImPress-P with PARA for the attack pattern.}
    \vspace{-0.05 in}
    \label{fig:paraattack}
\end{figure}

Let us first consider the case of Rowhammer (K=0).  For each activation, PARA issues a mitigation with probability {\em p}.  Each mitigation performs 4 activations (two victims on each side of the aggressor row).  Thus, the overhead of PARA is 4p per activation. For TRH of 4000/2000/1000, the value of p equals 1/84, 1/42, and 1/21. At p=1/84, the mitigation overhead of PARA is 4.76\%, as shown in Figure~\ref{fig:paraattack}.

For analyzing Row-Press, we vary K. The time for each iteration is (K+1)*tRC. With ImPress-P, the mitigation probability of PARA for each iteration of the loop would increase in proportion to (K+1), so it becomes (K+1)*p.  The probability can reach a maximum value of 1, so the effective mitigation probability for each loop would be MIN(1, p*(K+1)). Thus, the mitigation overhead of PARA would be 4*MIN(1, p*(K+1))*tRC per (K+1)*tRC, as shown in Equation~\ref{eq:paraattack}.

\begin{equation}   
\label{eq:paraattack}
    Slowdown = \frac{4*MIN(1, p \cdot (K+1))}{(K+1)}  
\end{equation}

Figure~\ref{fig:paraattack} shows the slowdown of PARA for TRH of 1000/2000/4000 as the amount of Row-Press (K) is varied from 0 to 100.  We note that Rowhammer is still the most potent attack.  The slowdown of Row-Press remains similar to Rowhammer until a critical point, after which  the slowdown starts to reduce (because the loop becomes large, PARA probability saturates at 1). Note that PARA has high mitigation overhead for both Rowhammer and Row-Press.

\section*{Acknowledgements}

We thank Salman Qazi for feedback on an earlier draft of the paper.  We also thank the anonymous reviewers of MICRO-2024 for their comments and suggestions.  

\bibliographystyle{IEEEtranS}
\bibliography{refs}

\end{document}